\documentclass[lettersize,journal]{IEEEtran}
\usepackage[compact]{titlesec}
\usepackage{amsmath,amsfonts}
\usepackage{algpseudocode}
\usepackage{algorithm}
\usepackage{array}
\usepackage[caption=false,font=footnotesize,labelfont=rm,textfont=rm]{subfig}
\usepackage{textcomp}
\usepackage{stfloats}
\usepackage{url}
\usepackage{verbatim}
\usepackage{graphicx}
\usepackage{cite}
\usepackage{multirow}
\usepackage{multicol}
\usepackage{booktabs}
\usepackage{paralist}
\newcommand{\myparagraph}[1]{\smallskip\textbf{#1.}}
\usepackage{makecell}
\usepackage{xcolor}

\usepackage{pifont} 
\usepackage{caption}
\newcommand{\xxx}{Intra-DP}
\newcommand{\xxxparallel}{LOP}
\newcommand{\xxxschedule}{LOSS}
\newcommand{\github}{\url{https://github.com/hku-systems/intraDP}}

\begin{document}

\title{\xxx: A High Performance Collaborative Inference System for Mobile Edge Computing}


\author{Zekai~Sun, Xiuxian~Guan, Zheng~Lin, Zihan~Fang, Xiangming~Cai,~\IEEEmembership{Member,~IEEE}, Zhe~Chen,~\IEEEmembership{Member,~IEEE}, Fangming~Liu,~\IEEEmembership{Senior Member,~IEEE,} Heming~Cui,~\IEEEmembership{Member,~IEEE}, Jie~Xiong,~\IEEEmembership{Senior Member,~IEEE}, Wei~Ni,~\IEEEmembership{Fellow,~IEEE}, and Chau~Yuen,~\IEEEmembership{Fellow,~IEEE}  
\thanks{Z. Sun, X. Guan, and H. Cui are with the Department of Computer Science, University of Hong Kong, Pok Fu Lam, Hong Kong SAR, China.  Z. Sun and H. Cui are also with Shanghai AI Laboratory, Shanghai, China. (e-mail: zksun@cs.hku.hk;  xxguan@cs.hku.hk; heming@cs.hku.hk).}
\thanks{Z. Lin, Z. Fang, and Z. Chen are with the
 Institute of Space Internet, Fudan University, Shanghai 200438, China, and also
 with the School of Computer Science, Fudan University, Shanghai 200438,
 China (e-mail: zlin20@fudan.edu.cn; zhfang19@fudan.edu.cn; zhechen@fudan.edu.cn).}
\thanks{X. Cai is with the School of Physics and
Information Engineering, Fuzhou University, Fuzhou 350108, China (e-mail:
 xiangming.cai@fzu.edu.cn).}
\thanks{F. Liu is with the Peng Cheng Laboratory, Shenzhen, China, and Huazhong University of Science and Technology, Wuhan, China (e-mail: fmliu@hust.edu.cn).}
\thanks{J. Xiong is with the School of Computing and Data Science, Nanyang Technological University, Singapore 639798 (e-mail: jie.xiong@ntu.edu.sg).}
\thanks{W. Ni is with Data61, CSIRO, Marsfield, NSW 2122, Australia, and the School of Computing Science and Engineering, and the University of New
 South Wales, Kennington, NSW 2052, Australia (e-mail: wei.ni@ieee.org).}
\thanks{C. Yuen is with the School of Electrical and Electronic Engineering, Nanyang Technological University, Singapore 639798 (e-mail:
 chau.yuen@ntu.edu.sg).}

}

\markboth{}%
{Shell \MakeLowercase{\textit{et al.}}: A Sample Article Using IEEEtran.cls for IEEE Journals}


\maketitle

\begin{abstract}
Deploying deep neural networks (DNNs) on resource-constrained mobile devices presents significant challenges, particularly in achieving real-time performance while simultaneously coping with limited computational resources and battery life.
While Mobile Edge Computing~(MEC) offers collaborative inference with GPU servers as a promising solution, existing approaches primarily rely on layer-wise model partitioning and undergo significant transmission bottlenecks caused by the sequential execution of DNN operations. 
To address this challenge, we present \xxx, a high-performance collaborative inference system optimized for DNN inference on MEC. 
\xxx{} employs a novel parallel computing technique based on local operators (i.e., operators whose minimum unit input is not the entire input tensor, such as the convolution kernel).
By decomposing their computations (operations) into several independent sub-operations and overlapping the computation and transmission of different sub-operations through parallel execution, \xxx{} mitigates transmission bottlenecks in MEC, achieving fast and energy-efficient inference.
The evaluation demonstrates that \xxx{} reduces per-inference latency by up to 50\% and energy consumption by up to 75\% compared to state-of-the-art baselines, without sacrificing accuracy.
\end{abstract}


\section{Introduction}
\label{sec:intro}

Deep Neural Networks~(DNNs) have empowered a wide range of mobile applications, from intelligent wearable sensors~\cite{luo2021binarized,tang2024merit} and autonomous vehicles~\cite{saridena2022dnn,lin2022channel,fang2024ic3m} to IoT systems~\cite{lin2024fedsn,zhao2019novel,peng2025sigchord,yuan2025constructing,chen2021rf,lin2021spatial,zhang2024fedac,yuan2024satsense,peng2024sums,zhao2024leo}. 
These applications are built upon advances in object detection~\cite{kapao,lin2024efficient}, robotic control~\cite{agrnav,lin2023pushing}, and environmental perception~\cite{cao2022monoscene}, all of which demand fast inference for swift responses. 
Deploying DNN models on real-world mobile devices (e.g., smartphones and IoT devices) presents two major challenges: the need for fast inference and the limitation of computing resources~\cite{lyu2023optimal,lin2024splitlora,zhang2024satfed,lin2024adaptsfl}, including computational power and battery life.

\begin{figure}[t!]
\setlength\abovecaptionskip{6pt}
\setlength\belowcaptionskip{-5pt}
\centering
\includegraphics[width=0.98\linewidth]{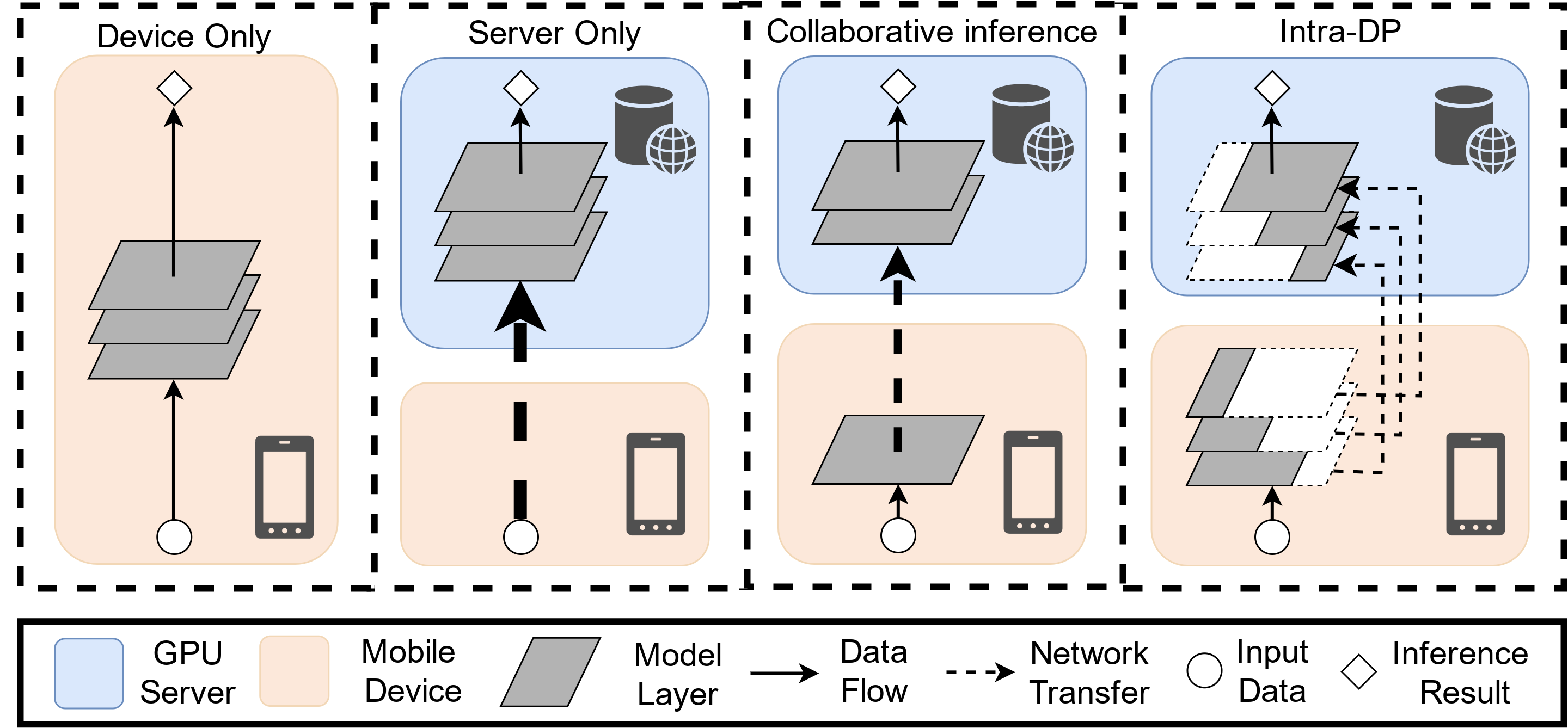}
\caption{\small Comparison between conventional Mobile Edge Computing approaches and \xxx{}. Inference results computed on the GPU server must be transmitted back to the mobile device for application utilization.}
\label{fig:scenario}
\end{figure}

Recent studies~\cite{abbas2017mobile, mohammed2020distributed, liang2023dnn, banitalebi2021auto, huang2020clio, hu2019dynamic, yang2021offloading, lin2025hasfl,kang2017neurosurgeon, wu2019efficient, chen2021energy} have proposed mobile edge computing (MEC) as a promising solution for fast and energy-efficient inference by leveraging GPU servers at the edge of the radio access network. 
Conventional MEC approaches (illustrated in Fig.~\ref{fig:scenario}) are categorized into device-only, server-only, and collaborative inference~\cite{liang2023dnn, chen2021energy}. 
Device-only inference is constrained by limited on-device computing resources (see Sec.~\ref{sec:background-device}), while server-only inference suffers from bandwidth limitations (see Sec.~\ref{sec:background-server}). 
Collaborative inference strikes a better balance between computation and communication through layer-wise model partitioning (layer partitioning), which distributes DNN layers across mobile devices and GPU servers.
Since some intermediate layers produce significantly smaller activations than raw input data~\cite{hu2019dynamic}, this approach reduces transmission overhead, further optimizing resource utilization (see Sec.~\ref{sec:background-layer}).

Despite its advantages, the sequential nature of layer partitioning introduces significant transmission delays~\cite{lin2024split,lin2025hsplitlora}, making it a major bottleneck for collaborative inference in MEC. 
Existing methods enforce strictly sequential execution, consisting of early-layer computation on mobile devices, intermediate result transmission, and final inference on GPU servers.
Limited network bandwidth (due to MEC network limitations and practical instability; see Sec.~\ref{sec:background-server}) prolongs transmission (about half of inference time in our experiments), causing significant idle periods on mobile devices.
This not only increases overall inference time but also results in significant energy waste (around 40\%). 
Although pipeline execution~\cite{gao2024accelerating} attempts to alleviate transmission overheads by overlapping computation and communication across multiple requests, its primary benefit lies in improving throughput, as opposed to reducing the inference time of individual requests, which is a crucial requirement for real-time mobile applications.

To address these limitations, a parallel execution strategy that overlaps computation and communication within a single inference request offers a promising solution. 
Unlike traditional sequential execution, this approach allows computation and data transmission to proceed concurrently, significantly reducing idle time on mobile devices. 
This speeds up inference and improves energy efficiency, as idle-period energy consumption is primarily driven by core components such as the CPU, GPU, and memory modules~\cite{kim2003leakage} (95\% of total energy in our experiment), while wireless network interface cards contribute only 1.5\%.

Implementing this parallel computing technique poses three key challenges. 
\textbf{\textcircled{1}} Traditional parallel computing methods (data, tensor, and pipeline parallelism) work well in data centers but struggle in MEC due to batch size limits, high synchronization costs, and transmission delays (see Sec.~\ref{sec:other_parallel}), necessitating a finer-grained scheduling unit.  
\textbf{\textcircled{2}} Fine-grained parallelism is vulnerable to consistency issues, where the failure or error of any single unit (e.g., receiving incorrect input) can compromise the entire inference result.  
\textbf{\textcircled{3}} A new scheduling paradigm is essential to balance computation and transmission within such fine-grained parallelism, yet its optimization is significantly complicated by two primary factors: an expanded search space from $O(n)$ (layers) to $O(n^2)$ (finer-grained scheduling units within layers), and the distinctive transmission and computing constraints inherent to MEC.

To address these challenges, we propose \xxx{} (Intra-Data Parallel), a high-performance collaborative inference system optimized for MEC. 
\textbf{\textcircled{1}} \xxx{} defines scheduling units based on local operators, where the minimum input is a subset of the full tensor (e.g., individual elements for ReLU~\cite{he2018relu} and tensor blocks for convolution~\cite{mohammed2020distributed}) and their computations (operations) can be decomposed into independent sub-operations (local operations). 
This intrinsic property allows local operations in subsequent layers to commence execution once their specific inputs are available, eliminating the need to await completion of all local operations in the current layer. 
Such a capability facilitates finer-grained parallel intra-data scheduling, thereby achieving computation-communication overlap even within a single inference request.
\textbf{\textcircled{2}} To ensure result correctness and completeness, we introduce Local Operation Parallelism (\xxxparallel), a novel parallel computing technique operating at the granularity of local operations. 
It guarantees that each operation receives the correct input and propagates the correct output during parallel execution by managing data dependencies, wherein the output of one operation serves as the designated input for another.
\textbf{\textcircled{3}} To achieve fast and energy-efficient inference, we propose the Local Operation Scheduling Strategy (\xxxschedule) for optimally distributing local operations between mobile devices and GPU servers. 
This strategy formulates the distribution task as a constrained optimization problem that leverages \xxxparallel, incorporates inherent MEC constraints, and is solved efficiently using differential evolution~\cite{qin2008differential}.

In summary, this paper makes the following contributions:
\begin{itemize}
    \item We propose \xxx, a novel collaborative inference system that enables fine-grained parallel scheduling based on local operators, specifically optimized for MEC. The code is released on \github. 
    \item We develop a comprehensive parallel computing technique (\xxxparallel) that guarantees inference correctness based on local operations while enabling efficient computation-transmission overlapping.
    \item We design an innovative scheduling algorithm (\xxxschedule) that achieves fast and energy-efficient inference based on a constrained optimization formulation for distributing local operations in MEC.
    \item We implement an adaptive control mechanism to dynamically handle real-time network bandwidth fluctuations in MEC.
    \item We empirically evaluate \xxx{} through extensive experiments, demonstrating its superiority over state-of-the-art baselines in MEC in terms of inference latency and energy efficiency. 
\end{itemize}

The rest of the paper is organized as follows. 
Sec.~\ref{sec:background} introduces the fundamental concepts and technical background of \xxx.
Sec.~\ref{sec:overview} presents the overview of \xxx{}, with detailed design in Sec.~\ref{sec:design}. 
Sec.~\ref{sec:implement} introduces the system implementation, followed by performance evaluation in Sec.~\ref{sec:evaluation}. 
Related works and technical limitations are discussed in Sec.~\ref{sec:discussion}.
Finally, conclusions are presented in Sec.~\ref{sec:conclusion}.

\section{Background and Motivation}
\label{sec:background}

\subsection{Device-only Inference}
\label{sec:background-device}
\begin{figure}[htp]
\vspace{-15pt}
\setlength\abovecaptionskip{6pt}
\setlength\belowcaptionskip{-5pt}
\centering
\subfloat[Inference Time\label{fig:device-inference}]{\includegraphics[width=0.48\linewidth]{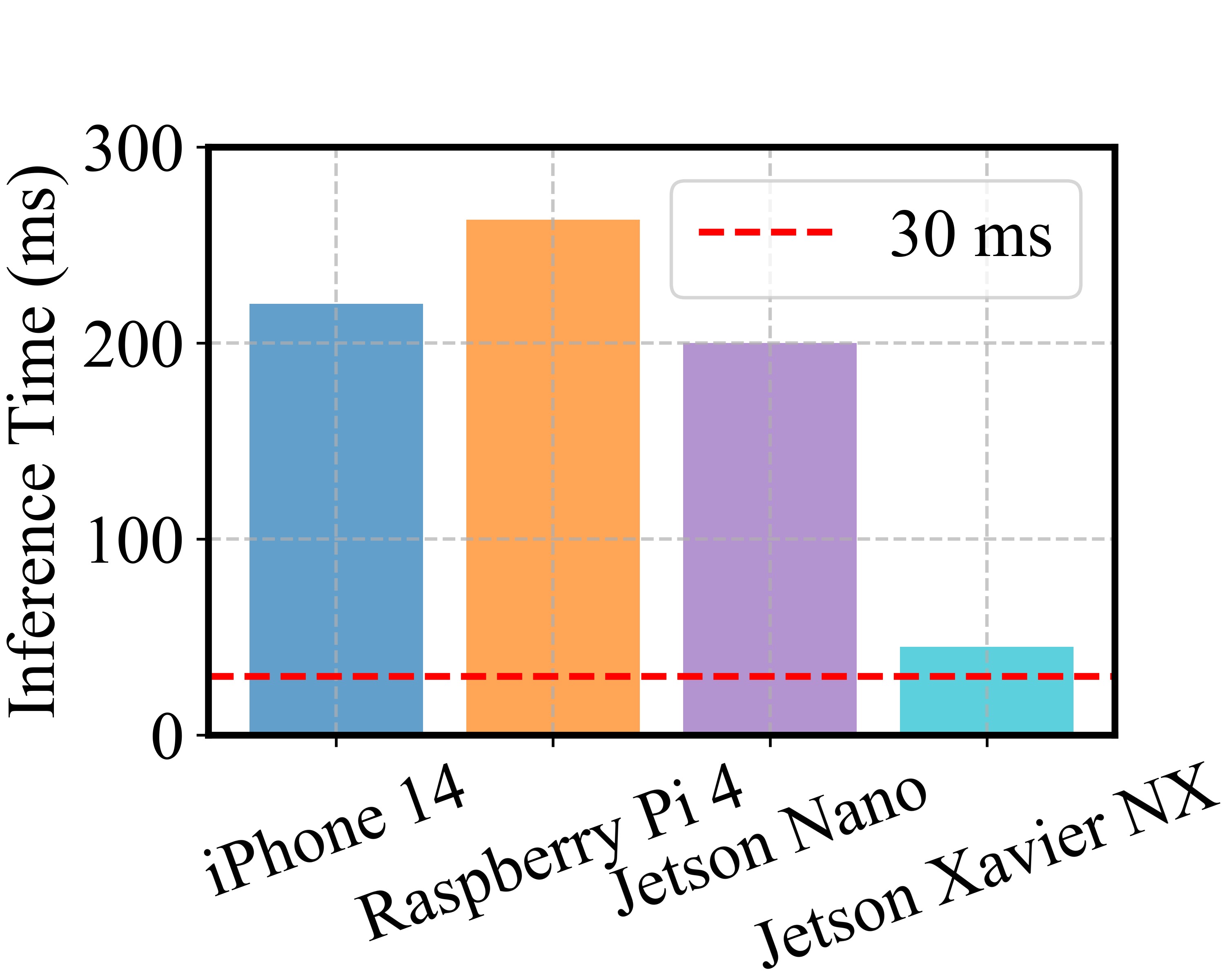}}
\hfil
\subfloat[Battery Life\label{fig:device-energy}]{\includegraphics[width=0.48\linewidth]{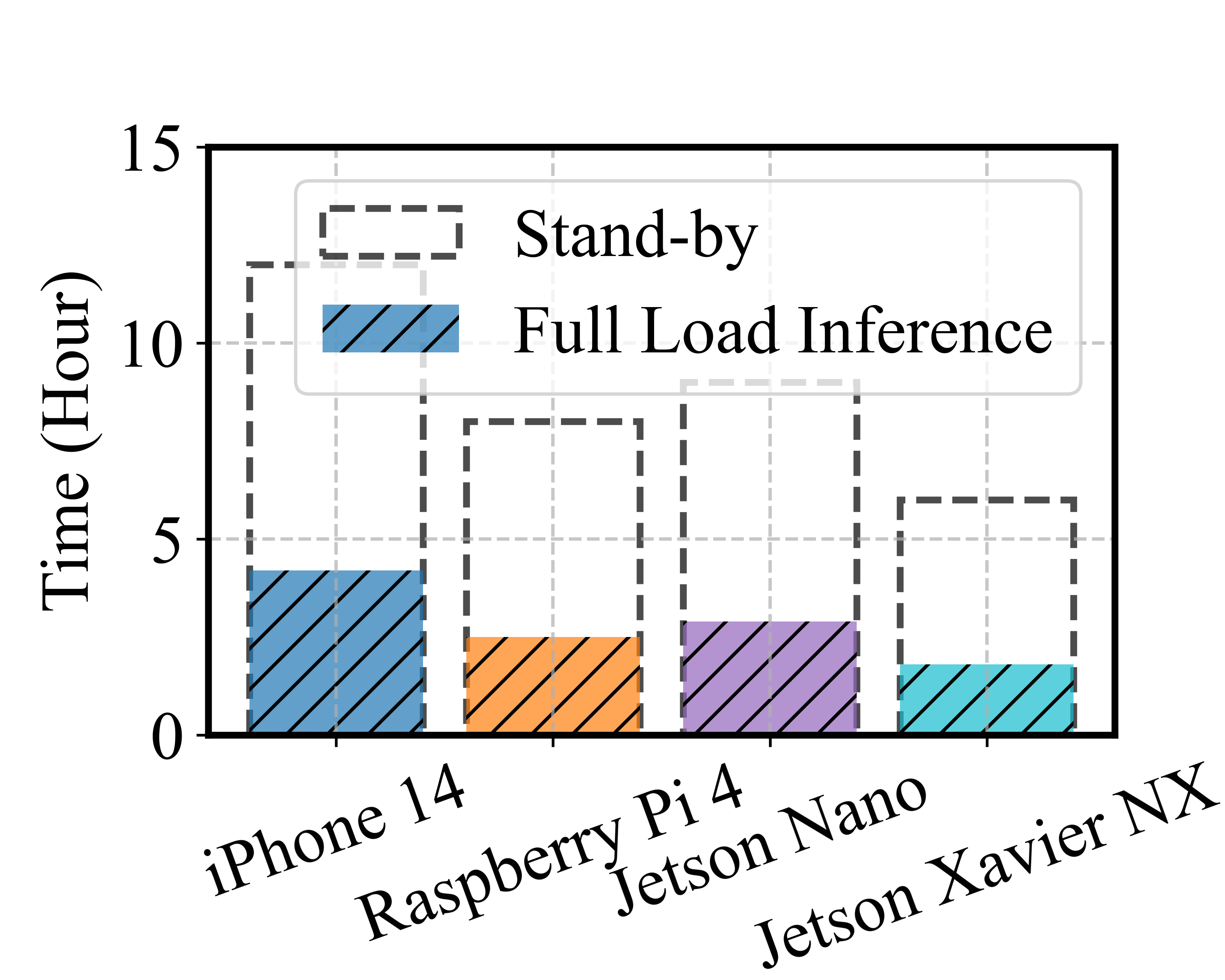}}
\caption{\small The performance of VGG-16 under device-only inference across different mobile devices~\cite{iphone, raspberrypi, jetsonnx, AnandTech, ning2024power}.}
\end{figure}

Device-only inference, which deploys DNN models directly on mobile devices, is severely constrained by the limited computational power of processors and battery capacity. 
As shown in Fig.~\ref{fig:device-inference}, the inference latency on these devices exceeds the 30 ms threshold required for smooth video fluency~\cite{ghosh2023react} (indicated by the red dotted line). 
Fig.~\ref{fig:device-energy} demonstrates that frequent model inference significantly reduces device standby time to 20–40\% of their normal duration, severely affecting user experience and overall device practicality.

\subsection{Server-only Inference}
\begin{figure}[t!]
    \centering
    \begin{minipage}[t]{0.48\linewidth}
        \centering
        \setlength\abovecaptionskip{0pt}
        \setlength\belowcaptionskip{0pt}
        \includegraphics[width=\linewidth]{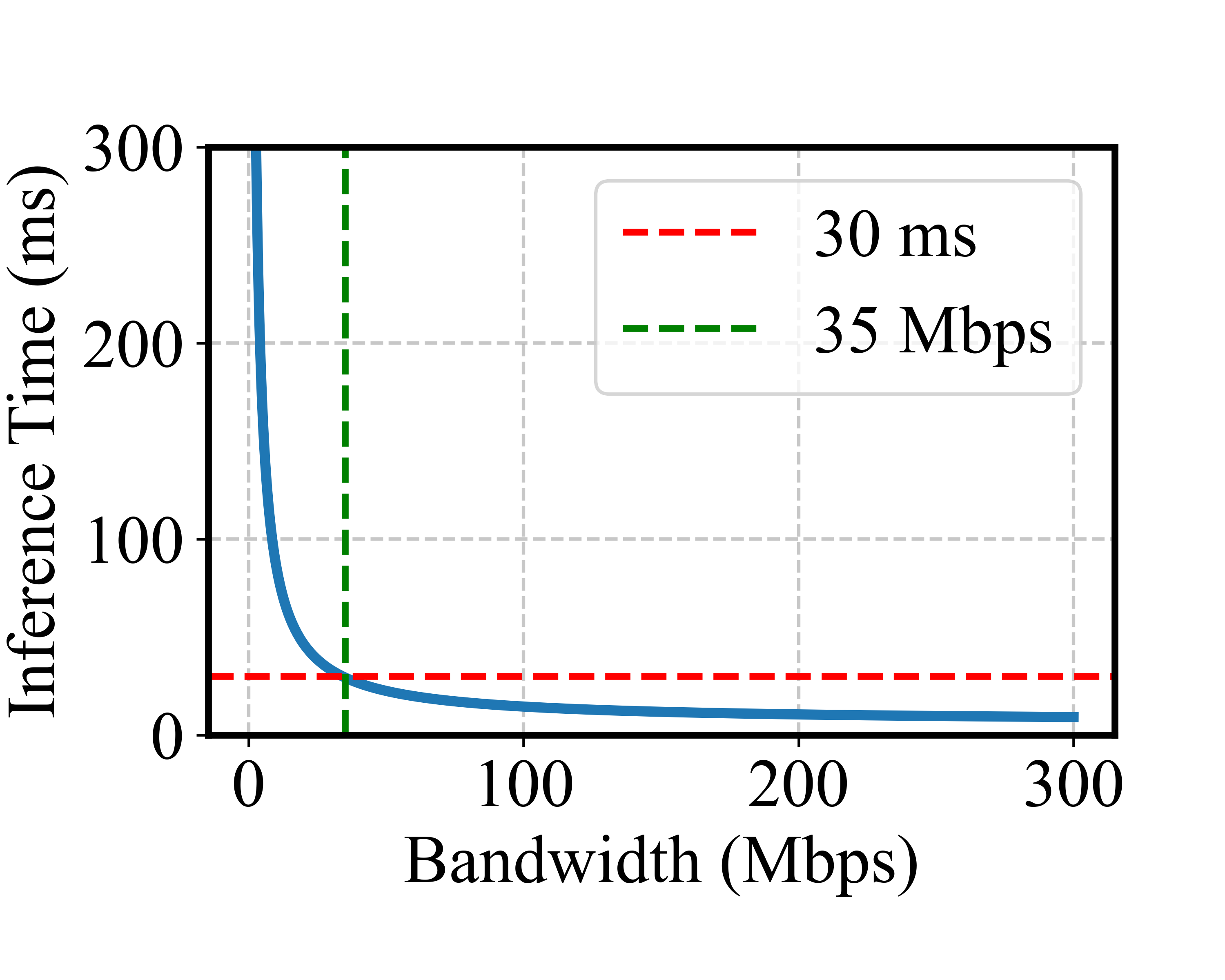}
        \caption{\small The inference time of VGG-16 under server-only inference across different network bandwidth.}
        \label{fig:server-inference}
    \end{minipage}
    \hfill
    \begin{minipage}[t]{0.48\linewidth}
        \centering
        \setlength\abovecaptionskip{0pt}
        \setlength\belowcaptionskip{0pt}
        \includegraphics[width=\linewidth]{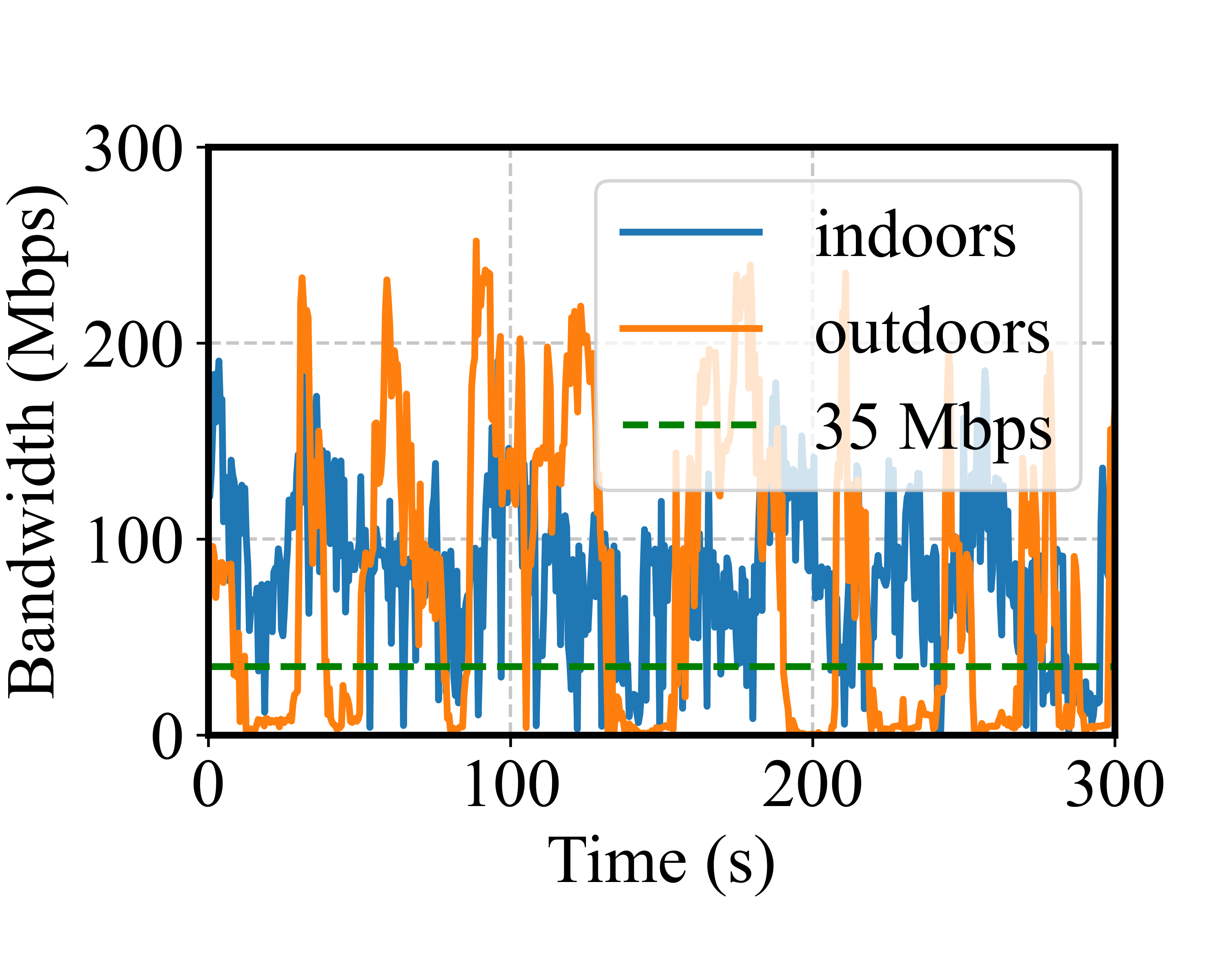}
        \caption{\small The wireless transmission instability of TCP between our robot and the base station in MEC networks.}
        \label{fig:bandwidth}
    \end{minipage}
\end{figure}

Server-only inference, which deploys DNN models directly on a GPU server, is susceptible to long tail delays
 and requires substantial bandwidth for data transmission. 
As shown in Fig.~\ref{fig:server-inference}, under the same testbed settings as Sec.~\ref{sec:implement} (a robot equipped with a Jetson Xavier NX~\cite{jetsonnx} and a GPU server with an NVIDIA GeForce GTX 3080), inference time increases sharply as bandwidth decreases. 
However, in real-world scenarios, mobile devices primarily rely on wireless networks, which offer high mobility but limited bandwidth compared to data center networks equipped with high-speed technologies (e.g., 40–500 Gbps for InfiniBand~\cite{infiniBand} and PCIe~\cite{li2019evaluating}).

\label{sec:background-server}

The inherent bandwidth capacity of wireless networks in MEC is subject to restrictions stemming from both theoretical limits and practical implementation factors.
While Wi-Fi 6 can achieve a peak bandwidth of 1.2 Gbps per stream~\cite{liu2023first}, mobile devices lack the necessary hardware to fully leverage this capacity~\cite{yang2022mobile}. 
The actual available bandwidth varies significantly due to factors such as device mobility~\cite{masiukiewicz2019throughput}, signal obstruction~\cite{ding2015performance}, and channel contention~\cite{ren2018proportional}.

To examine wireless instability in MEC scenarios, we conducted a robot surveillance experiment where four-wheeled robots navigated through a lab (indoors) and a campus garden (outdoors) at speeds of 5–40 cm/s. 
Using iperf~\cite{iperf}, we measured real-time wireless bandwidth capacity between the robot and a base station over TCP~\cite{tian2005tcp} at 0.1-second intervals for five minutes. 
As shown in Fig.~\ref{fig:bandwidth}, the average bandwidth was 93 Mbps indoors and 73 Mbps outdoors, with outdoor measurements exhibiting higher fluctuations and occasional near-zero drops due to obstacles and reduced signal reflections, confirming that server-only inference is difficult to sustain under realistic wireless network conditions.


\subsection{Related Parallel Computing Techniques}
\label{sec:other_parallel}

Parallel computing techniques have been extensively explored and proven effective in modern data centers~\cite{narayanan2021efficient, zhuang2023optimizing, gao2024accelerating}. 
However, they are unsuitable for MEC due to fundamental differences in computational constraints and latency requirements.
The three primary parallelism strategies in data centers are data parallelism (DP), which replicates the model across devices and processes mini-batches in parallel; tensor parallelism (TP), which partitions computations within a layer across multiple devices; and pipeline parallelism (PP), which distributes different layers of DNN model across devices (layer partitioning) and executes them sequentially (pipeline execution) to reduce idle time.

DP's efficiency is fundamentally constrained by batch size requirements~\cite{narayanan2021efficient}.  
In data centers, large batch sizes (e.g., 16 images) are split into mini-batches (e.g., 2 images each), maintaining computational efficiency.  
However, MEC requires real-time inference with minimal latency, typically processing a single input at a time (e.g., 1 image). 
Further splitting such small inputs into sub-mini-batches (e.g., 1/4 of an image) is impractical, making DP inefficient.

\begin{figure}[!t]
\vspace{-10pt}
\setlength\abovecaptionskip{6pt}
\setlength\belowcaptionskip{0pt}
    \centering
    \subfloat[Inference Time]{\includegraphics[width=0.48\linewidth]{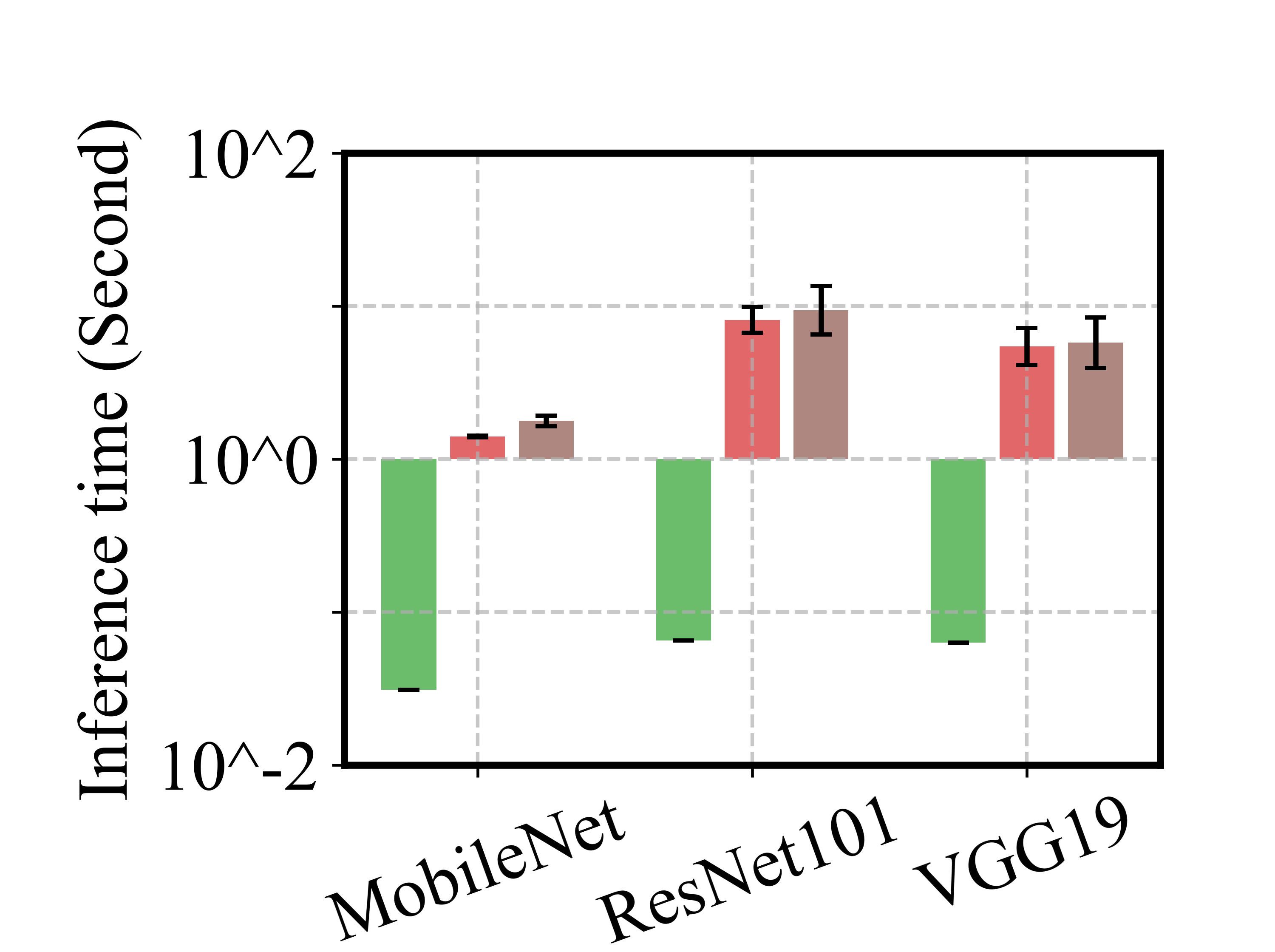}}
    \hfil
    \subfloat[Energy Consumption]{\includegraphics[width=0.48\linewidth]{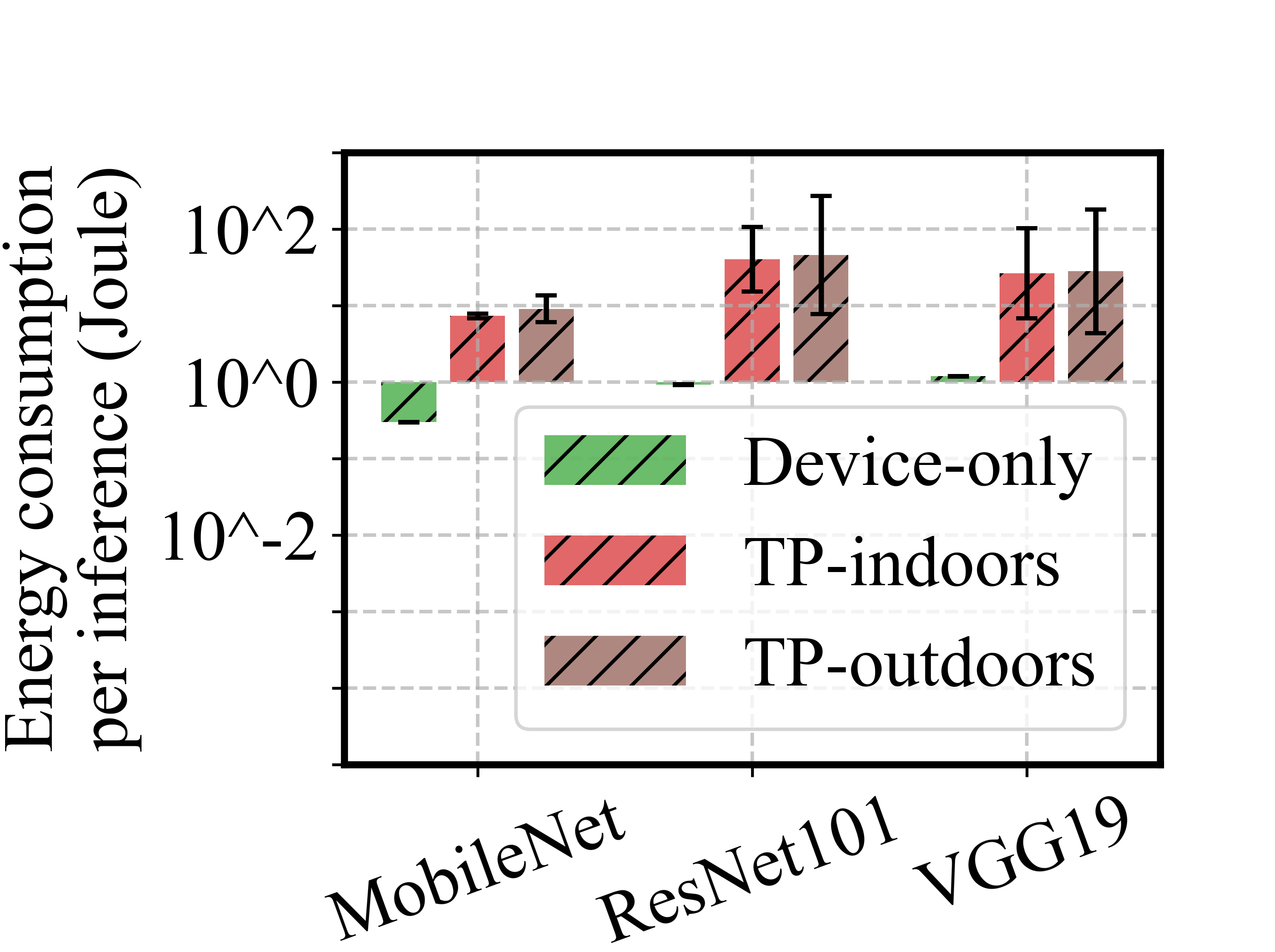}}
    \caption{\small The performance of TP for different models.}
    \label{fig:tp_end2end} 
\end{figure}

TP enables parallel computation within a single layer across devices but incurs substantial communication overhead due to the required all-reduce synchronization~\cite{zhuang2023optimizing}. 
We evaluated DINA~\cite{mohammed2020distributed}, a state-of-the-art TP method, on our testbed (see Sec.\ref{sec:implement}). 
As shown in Fig.~\ref{fig:tp_end2end}, the excessive all-reduce overhead results in an inference time 45.2 to 143.9 times, and an energy consumption per inference 28.5 to 62.7 times, that of device-only inference.
While there have been efforts to reduce the communication overhead of TP in data centers, such as distributing each layer along both the spatial and temporal dimensions~\cite{wang2024primepar}, these methods remain ineffective in MEC due to the persistent all-reduce requirement. 
In contrast, \xxx{} eliminates this communication for local operators, further reducing overhead.

PP employs pipeline execution to enhance throughput by overlapping computation and communication across multiple requests. However, this approach does not reduce the completion time of a single inference~\cite{gao2024accelerating}.  
\cite{laskaridis2020spinn_mobicom} attempts to optimize transmission efficiency through an early-exit policy combined with layer partitioning. 
While this reduces the number of transmissions at the cost of accuracy, it introduces unpredictable and incorrect inference results, and requires careful expert tuning of parameters. 
Additionally, transmission delays persist in \cite{laskaridis2020spinn_mobicom}, an issue that can be effectively addressed by the combination of \xxx{} without sacrificing accuracy.

In conclusion, while conventional parallel computing techniques have proven effective in data centers, they are unsuitable for MEC. 
DP is limited by batch size constraints, TP suffers from excessive synchronization overhead, and PP’s pipeline execution focuses solely on throughput, offering no provision for inference latency reduction.
Consequently, existing collaborative inference in MEC relies on layer partitioning, which enforces sequential execution and introduces transmission delays, making them a major bottleneck for low-latency inference.

\subsection{Existing Collaborative Inference}
\label{sec:background-layer}

\begin{figure}[t!]
\vspace{-10pt}
\setlength\abovecaptionskip{-10pt}
\setlength\belowcaptionskip{0pt}
\centering
\includegraphics[width=0.82\linewidth]{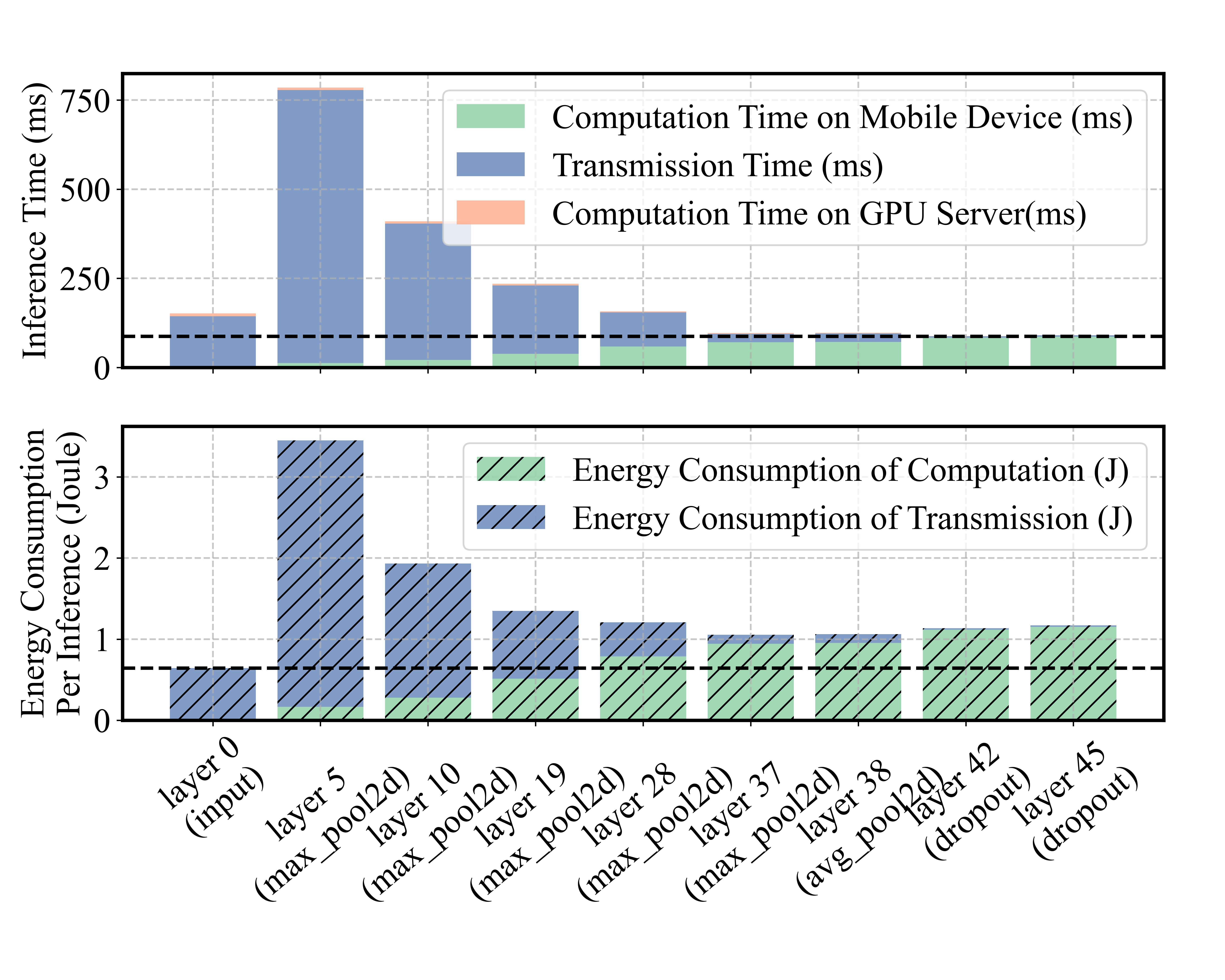}
\caption{\small The performance of different layer partitioning strategies for VGG-19 under 35 Mbps in our experiments. 
    The X-axis represents various partitioning points, where 'layer i' denotes that all layers up to and including the $i_{th}$ layer are executed on the robot, while the remaining layers are offloaded to the GPU server.}
\label{fig:layer_partitioning}
\end{figure} 

Existing collaborative inference methods~\cite{liang2023dnn, banitalebi2021auto, huang2020clio, hu2019dynamic, yang2021offloading, kang2017neurosurgeon, wu2019efficient, chen2021energy} primarily focus on layer partitioning to balance inference time and energy consumption, as shown in Fig.~\ref{fig:layer_partitioning}.
``Layer 0'' represents server-only inference, where all model layers run on a GPU server, and transmission time varies with network bandwidth, even under the same partitioning strategy.
These methods can be categorized by their optimization objectives: accelerating DNN inference\cite{hu2019dynamic, banitalebi2021auto, huang2020clio, yang2021offloading, kang2017neurosurgeon, liang2023dnn} or minimizing energy consumption under deadline constraints~\cite{wu2019efficient, chen2021energy}.
The widespread adoption of layer partitioning in mobile applications has drawn significant research attention~\cite{banitalebi2021auto, huang2020clio, hu2019dynamic, yang2021offloading, kang2017neurosurgeon, liang2023dnn}, as optimal strategies depend on multiple factors, including model architecture, hardware capabilities (e.g., GPU servers and mobile devices), network conditions, and application-specific trade-offs between inference speed and energy efficiency.
However, all existing methods~\cite{liang2023dnn, chen2021energy, banitalebi2021auto, huang2020clio, hu2019dynamic, kang2017neurosurgeon, wu2019efficient, yang2021offloading} are inherently constrained by transmission delays due to the sequential nature of layer partitioning, a limitation that can be effectively addressed by \xxx{}.

While some methods attempt to mitigate transmission delays in collaborative inference through finer-grained scheduling units, such approaches lose effectiveness in MEC scenarios. 
\cite{9353250, zeng2021coedge} propose input-splitting methods, distributing raw input across devices based on their computational power. 
However, their design for IoT device groups is ill-suited for MEC; the substantial computational disparity between mobile devices and GPU servers therein leads to near-complete reliance on server-side execution, effectively reverting to server-only inference.
\cite{bin2024coacto} employed tile-based scheduling for concurrent computation and communication, but the resulting input fragmentation from small tiles leads to substantial overhead of frequent invocation. 
To mitigate this, they utilize an asynchronous execution approach, which is CPU-restricted and consequently precludes effective GPU acceleration in MEC. 
Separately, their selection of coarser tile granularity, coupled with a simplistic, unscheduled greedy strategy, impairs overall parallel execution efficiency. 
These distinct limitations underscore the necessity for \xxx.

\section{System Overview}
\label{sec:overview}
In this section, we introduce \xxx, a high-performance collaborative inference system optimized for MEC.
In this setting, resource-constrained mobile devices (e.g., robots, drones, and smartphones) collaborate with GPU servers through wireless networks (e.g., Wi-Fi 6/5G) to achieve real-time and energy-efficient inference on local inputs. 

\subsection{Key Insight}
Existing methods are limited to layer partitioning, which enforces serial execution and limits parallelism. 
Each DNN layer comprises one or more operators (e.g., convolution~\cite{mohammed2020distributed}, Softmax~\cite{liu2016large}), which must be executed sequentially to ensure computational correctness. 
As a result, current approaches can only partition these serial execution sequences at the layer level, restricting opportunities for parallel execution.  

Our key insight is that, although operators are the smallest computation units at the system level, their computations (operations) can still be decomposed into several independent sub-operations based on their computational properties. 
Specifically, local operators, whose computations do not require access to the entire input tensor, enable finer-grained parallel execution. 
Existing methods treat an operator’s computation based on its entire input as a single, indivisible operation. 
In contrast, \xxx{} leverages the inherent divisibility within local operators by treating each individual computation, based on its minimal input unit within the local operator, as an independent local operation.
Since these local operations are independent, those local operations in subsequent layers can be computed as soon as their required inputs are ready, without waiting for all operations in the current layer to finish. 
This independence enables a new parallelization opportunity within serial execution sequences, breaking the strict sequential execution constraint.



\subsection{Overall Workflow}
\begin{figure}[!t]
\setlength\abovecaptionskip{6pt}
\setlength\belowcaptionskip{0pt}
    \centering
    \includegraphics[width=0.88\linewidth]{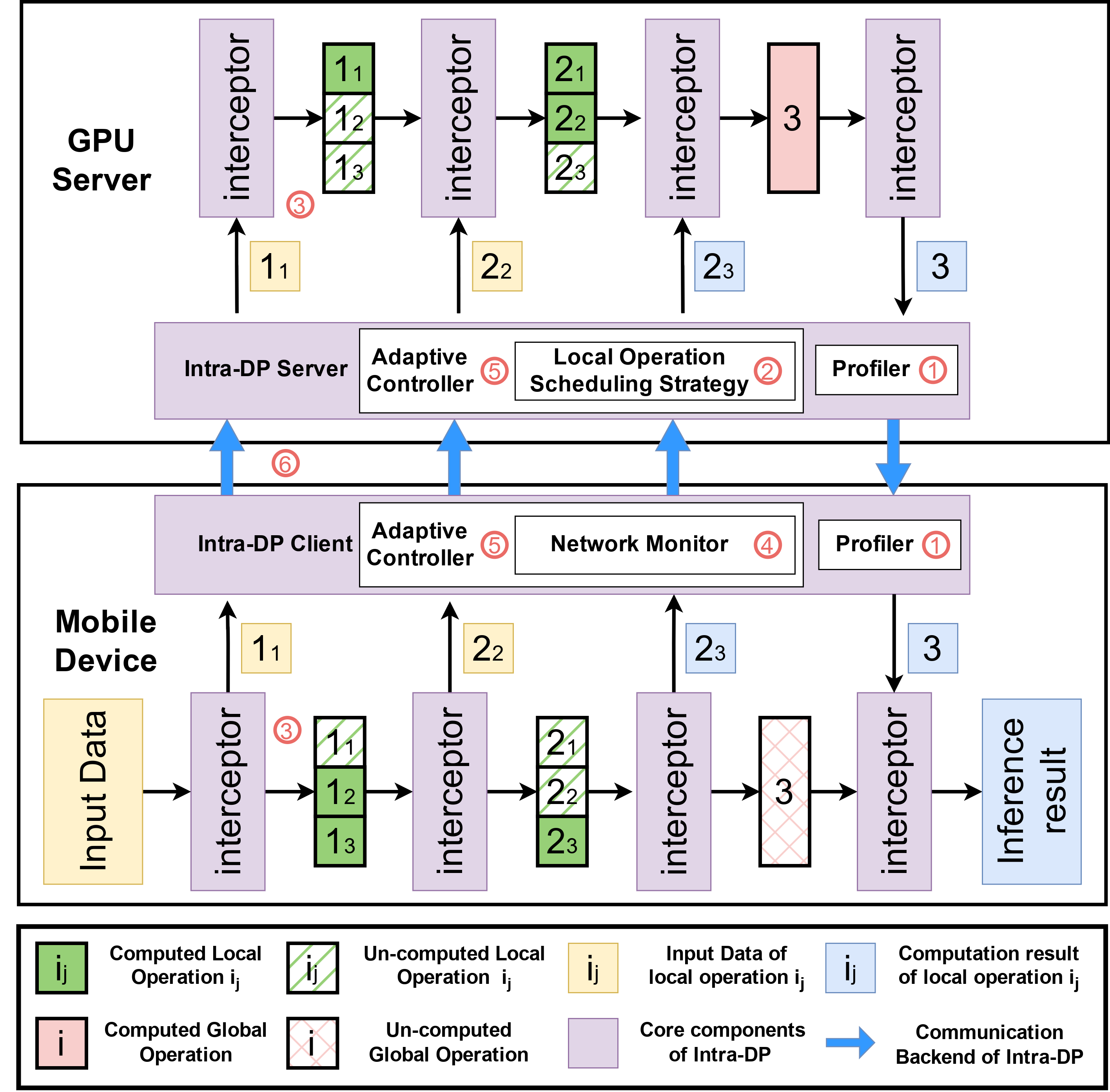}
    \caption{\small Overview and inference workflow of \xxx.
    Local operation \( i_j \) represents the \( j \)-th local operation within the \( i \)-th operator of the model.
    }
    \label{fig:architecture}
\end{figure}

As illustrated in Fig.~\ref{fig:architecture}, \xxx{} consists of three stages: Profiling, Scheduling Optimization, and Runtime.

The first stage, Profiling (\textbf{\textcircled{1}}), is an offline process that collects essential runtime traces for Scheduling Optimization. 
It runs only once and captures key execution characteristics, including: \begin{inparaenum}[i)] \item  operator types and input/output sizes; \item  execution times of operations on the mobile device and GPU server; \item data dependencies between operations (i.e., the output of one operation serves as the input for another).\end{inparaenum}

Using this profiling data, the system enters the Scheduling Optimization stage, where it applies Local Operation Scheduling Strategy (see Sec.~\ref{sec:schedule}, \textbf{\textcircled{2}}) to achieve low-latency, energy-efficient inference. 
This strategy determines the optimal distribution of local operations, balancing computational workload and transmission synchronization. 
To mitigate the impact of network fluctuations, \xxx{} precomputes optimal scheduling plans for varying network bandwidth conditions in 1 MB intervals within the typical MEC bandwidth range (0\textasciitilde50 MB), eliminating runtime optimization overhead.

During Runtime, \xxx{} utilizes Operator Interceptors (see Sec.~\ref{sec:operator}, \textbf{\textcircled{3}}) to manage input/output tensor partitioning and reconstruction for each operator, enabling partial tensor processing based on local operations. 
To ensure reliability in MEC, the system continuously monitors network conditions utilizing established tools~\cite{10.1145/1410077.1410081} (\textbf{\textcircled{4}}) and Adaptive Controllers (see Sec.~\ref{sec:adaptive}, \textbf{\textcircled{5}}) dynamically adjust scheduling strategies based on real-time bandwidth fluctuations, ensuring stable performance.  
Then, Local Operation Parallelism (see Sec.~\ref{sec:parallel}, \textbf{\textcircled{6}}) enables parallel execution between the client and server while preserving data dependencies for correct inference. 
To support seamless transitions between different schedules without delays, \xxx{} maintains a model replica on the GPU server (Fig.~\ref{fig:architecture}), allowing instant adaptation to network variations.

Despite these optimizations, \xxx{} introduces negligible overhead beyond the original inference process. 
The only additional cost stems from tensor splitting and combining.  
However, the impact of tensor splitting is effectively masked because it executes concurrently with data transmission and computation, allowing its associated latency to be absorbed by these parallel tasks.
While tensor combining introduces potential waiting time due to data dependencies, \xxxschedule{} minimizes this impact by formulating the waiting overhead as a nonlinear optimization problem and solves it efficiently. 
This careful design ensures that \xxx{} maintains high execution efficiency with minimal extra cost.

\section{Detailed Design}
\label{sec:design}
In this section, we detail the design of \xxx, a collaborative inference system enabling fine-grained parallel scheduling based on local operators, specifically optimized for MEC. 
We first identify and categorize local versus global operators (see Sec.~\ref{sec:operator}), which establishes the foundation for our parallel execution strategy. 
Building upon this, Sec.~\ref{sec:parallel} introduces \xxxparallel, a parallel computing technique that guarantees inference correctness with these local operations while achieving efficient computation-transmission overlap. 
To effectively orchestrate these parallel local operations, Sec.~\ref{sec:schedule} presents \xxxschedule, an innovative scheduling algorithm that achieves fast and energy-efficient inference by formulating the distribution of local operations as a constrained optimization problem. 
Finally, to ensure robust performance under MEC's inherent network fluctuations, Sec.~\ref{sec:adaptive} describes an adaptive control mechanism designed for dynamic real-time network bandwidth management.

\subsection{Local Operators and Global Operators}
\label{sec:operator}
As the performance gains of \xxx{} rely on the parallel execution of local operations, it is crucial to determine the minimal input unit required for each operator based on its computational characteristics. 
To achieve this, \xxx{} employs Operator Interceptors (Fig.~\ref{fig:architecture} \textbf{\textcircled{3}}) to handle input/output tensor partitioning and reconstruction, enabling partial tensor processing, and classify operators into local and global categories: local operators process only a subset of the input tensor, whereas global operators require the entire tensor. 
We identify three common types of local operators in models commonly used on mobile devices (mobile models):

\begin{itemize}
    \item Element-wise local operators operate on individual tensor elements and are widely used in activation functions (e.g., ReLU~\cite{he2018relu}, Sigmoid~\cite{ramapuram2024theory}, and SiLU~\cite{nishiyama2020silu}). 
    However, activation functions like Softmax~\cite{liu2016large} require access to the entire input tensor and are therefore global operators.
    

    \item Block-wise local operators process tensor blocks at corresponding positions, commonly found in convolution-related layers (e.g., convolution~\cite{mohammed2020distributed} and max pooling~\cite{sun2021ampnet}). 
    The block size is determined by the operator’s parameters (e.g., kernel size, padding, and dilation)~\cite{9353250}.
    

    \item Row-wise local operators operate on individual rows of the input tensor and are widely used in matrix operations (e.g., addition and multiplication). 
    This operational paradigm involves partitioning the input matrix by rows, enabling each device to process an independent subset while critically operating with a shared (or replicated, i.e., non-partitioned) layer parameter matrix. 
\begin{equation} 
\left(
\begin{smallmatrix}
a_{1} \\
\vdots \\
a_{m}
\end{smallmatrix}
\right)
\times
\left(
\begin{smallmatrix}
b_{1} & \cdots & b_{n}
\end{smallmatrix}
\right)
=
\left(
\begin{smallmatrix}
c_{11} & \cdots & c_{1n} \\
\vdots & \ddots & \vdots \\
c_{m1} & \cdots & c_{mn}
\end{smallmatrix}
\right)
\end{equation}
The capacity of these operators to eliminate all-reduce communication is founded on a core matrix calculation principle: an operation on an input row $a_{1}$ yields an output row $\left( c_{11} \cdots c_{1n}\right)$ that is itself directly processable by subsequent matrix operators in a similar row-wise manner. 
This self-contained, per-row processing characteristic inherently obviates the need for inter-device synchronization via all-reduce. 
TP, in contrast, partitions the layer's parameter matrix and replicates the full input matrix across devices, a strategy that necessitates all-reduce operations to aggregate partial results. 
Row-wise local operators circumvent this costly communication by enabling independent row processing on each device coupled with a shared parameter matrix.
Moreover, \xxxparallel{} treats operators with layer parameter matrices containing only one row as global operators.

\end{itemize}


\begin{table}[h]
\vspace{5pt}
\setlength\abovecaptionskip{6pt}
    \centering
    \setlength{\tabcolsep}{1pt}  
    \begin{tabular}{c c | c c}  
    \toprule
    Model & Local / Global & Model & Local / Global \\
    \midrule
    DenseNet~\cite{huang2018densely}  & 428 / 3  & RegNet~\cite{xu2022regnet}   & 231 / 3 \\
    ResNet~\cite{targ2016resnet}  & 341 / 3  & VGGNet~\cite{simonyan2015deep} & 73 / 5 \\
    ResNeXt~\cite{xie2017aggregatedresidualtransformationsdeep} & 342 / 3 & ConvNeXt~\cite{woo2023convnext} & 340 / 6 \\  
    \bottomrule
    \end{tabular}
    \caption{\small Number of local/global operators in mobile models.}
    \label{tab:local-and-global}
\end{table}

Models with many local operators, prevalent in mobile models (e.g., CNNs for computer vision~\cite{kapao} and point cloud processing~\cite{agrnav}), can benefit from \xxx{} through parallel execution.  
Table~\ref{tab:local-and-global} quantifies their distribution across several mobile models based on the above definition.


While local operators dominate CNN-based models, trans\-former-based models rely on self-attention mechanisms, which use Softmax~\cite{liu2016large}, a typical global operator, to compute attention weights over entire input sequences. 
Since each transformer layer contains at least one global operator, the performance gains of \xxx{} over existing collaborative inference methods are limited, as its advantages primarily come from parallelizing local operations, which is constrained by the presence of global operators. 
However, \xxx{} still outperforms existing methods by enabling parallel execution of local operations between consecutive global operators and seamlessly falling back to conventional layer partitioning when handling models without local operators.

\subsection{Local Operation Parallelism}
\label{sec:parallel}
In this section, we demonstrate how \xxx{} ensures inference correctness based on local operators and how \xxxparallel{} achieves low latency and energy efficiency through the parallel execution of local operations (Fig.~\ref{fig:architecture} \textbf{\textcircled{6}}).  

\xxxparallel{} guarantees inference correctness by comprehensively managing dependencies between local and global operators. 
It tracks data dependencies to ensure that the output of one operation is correctly used as the input for subsequent operations, considering a dependency established if any part of the output serves as input.
For local operators, it preserves computational correctness by enforcing the appropriate execution sequence and ensuring that each operation receives the correct input (either a raw input data or the output from a preceding operation) based on the DNN model structure and operator computation logic. 
For global operators, synchronization barriers are introduced to guarantee that all required input tensors are fully assembled before execution.
By maintaining proper data flow throughout the inference process, this approach guarantees correctness across both local and global operators.

\begin{figure}[!t]

\setlength\abovecaptionskip{6pt}
\setlength\belowcaptionskip{0pt}
    \centering
    \includegraphics[width=0.95\linewidth]{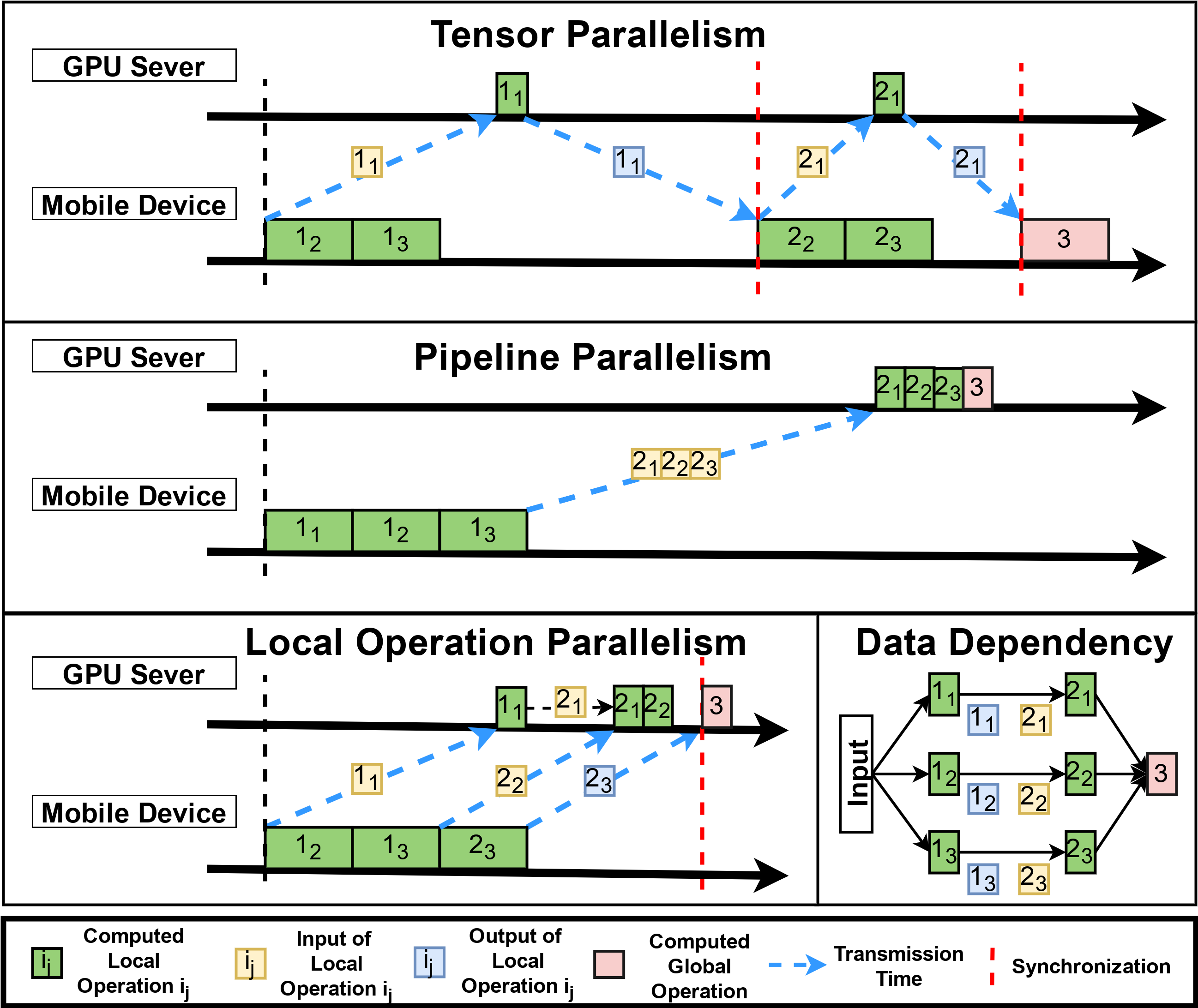}
    \caption{\small Workflow of TP, PP and \xxxparallel.
    In the three cases above, each local operator executes three operations with identical computation times on both the mobile device and the GPU server, along with the corresponding transmission time, while the lower right corner illustrates the data dependencies between operations.
    }
    \label{fig:overview}
\end{figure}

Fig.~\ref{fig:overview} compares the workflow of \xxxparallel{} with conventional TP and PP (layer partitioning).
To reduce transmission overhead, \xxxparallel{} employs two key optimizations. 
First, it overlaps computation and transmission for different local operations to maximize execution efficiency (e.g., transmitting the input for `LO$1_{1}$' while computing `LO$1_{2}$', `LO$1_{3}$', and `LO$2_{3}$' on the robot). 
Second, it exploits data locality by prioritizing execution on devices that already contain the required input (e.g., `LO$2_{1}$' directly using the output of `LO$1_{1}$' on the GPU server). 
Unlike TP, which suffers from high communication costs due to frequent all-reduce synchronization, \xxxparallel{} limits synchronization to global operators only, eliminating unnecessary data transfers. 
Although \xxxparallel{} incurs finer-grained communication and a higher volume than PP, its early overlapping strategy effectively reduces overall transmission time, outperforming PP’s sequential execution for individual inference requests.
Through the effective overlapping of computation and communication, \xxx{} curtails idle time on mobile devices, thereby accelerating inference and bolstering energy efficiency. 
This improvement is significant because energy consumption in idle states is chiefly governed by core components (e.g., CPU, GPU, and memory modules), as opposed to wireless network interface cards.

\subsection{Local Operation Scheduling Strategy}
\label{sec:schedule}
In this section, we detail how \xxx{} optimizes computation and transmission scheduling for local operations, achieving fast and energy-efficient inference (Fig.~\ref{fig:architecture} \textbf{\textcircled{2}}).  

To ensure efficient inference across varying network conditions, \xxxschedule{} operates under the assumption of stable bandwidth during each individual inference request. 
This assumption is justified because the typical model inference duration (tens to hundreds of milliseconds) is significantly shorter than the timescale of common wireless bandwidth fluctuations (on the order of seconds~\cite{9367147}). 
Leveraging this stability, \xxxschedule{} formulates the local operation scheduling problem for a given network bandwidth as a constrained optimization problem, which is then efficiently solved using differential evolution~\cite{qin2008differential}.
To streamline heterogeneous operation integration, it unifies global and local operations by treating global operators as special cases that aggregate all outputs from preceding local operations. 
The key performance boost stems from computation-communication overlapping, where \xxxschedule{} leverages data locality by prioritizing execution on devices that already contain the required input.

When multiple operations, particularly block-wise ones, located on both mobile devices and GPU servers concurrently depend on the same input, this shared input must be transferred irrespective of its original location. 
To further optimize such scenarios, \xxxschedule{} exploits data locality by introducing local operation replication, which allows select operations to be executed simultaneously on both mobile devices and GPU servers. 
The effectiveness of local operation replication stems from the common scenario where the shared input is the output of a preceding operation; replication ensures this output, now the required shared input, can be directly obtained locally on each device without extra transmission. 
This strategy incurs minimal computational redundancy but yields significant reductions in data transmission costs, proving particularly beneficial for distributed operations that exhibit shared input dependencies.

\subsubsection{DNN Execution Model}
A DNN is represented as a directed acyclic graph \( G = (V, E) \), where vertices \( V \) denote layers (operators) and edges \( E \) indicate data dependencies. 
Two virtual vertices,  \( input \) and \( output \), represent the model’s initial input and final inference result. 
The execution spans two instances: one on a resource-constrained mobile device (\( M \)) and the other on a GPU server (\( R \)). 


\subsubsection{Variables and Functions}
We treat each calculation of an operator based on its minimum input unit as an operation, and define \( OP(v) \) as the set of operations for vertex \( v \), where global operators always have \( |OP(v)| = 1 \). 
Operation allocations to \( M \) and \( R \) are represented by \( x_M(v) \), \( x_R(v) \subseteq OP(v) \), with overlaps (\( x_M(v) \cap x_R(v) \neq \emptyset \)) occurring when local operations are replicated, especially for block-wise operators.  

Execution starts at times \( s_M(v), s_R(v) \geq 0 \), and computation durations on each device are estimated as \( C_M(v) \) and \( C_R(v) \). 
If no operation is performed on a device, its computation time is zero.
Data transmission times include \( T_{\text{MR}}(u, v) \) for sending data from \( M \) to \( R \) and \( T_{\text{RM}}(u, v) \) for the reverse direction, both determined by network bandwidth and dependent on the volume of input/output data associated with the respective operations.

\subsubsection{Objective Function}

The primary objective of \xxxschedule{} is to minimize end-to-end inference latency while ensuring that the final results are available on the mobile device. 
This is achieved by minimizing $T$, defined as:
\begin{equation}
\min T = s_M(\text{output}),
\end{equation}
where $s_M(output)$ represents the time at which the mobile device receives the final inference result, thereby reducing the idle time on mobile device. 
While \xxx{} can be adapted to prioritize minimizing energy consumption under specified deadline constraints, similar to approaches in~\cite{wu2019efficient, chen2021energy}, by modifying this objective function, such exploration is not pursued here. 
Optimizing such application-specific trade-offs between inference speed and energy efficiency falls beyond the scope of this paper.

\subsubsection{Constraints}
\myparagraph{Data Partitioning}
To ensure correctness, all operations must be executed: $x_M(v) \cup x_R(v) = OP(v), \forall v \in V$.

\myparagraph{Location}
Input data is initially available on \( M \), ensuring \( x_M(input) = input \) and \( x_R(input) = \emptyset \). 
Similarly, the final results must reside on \( M \), so \( x_M(output) = output \) and \( x_R(output) = \emptyset \).

\myparagraph{Data Dependency}
Operations start only after receiving all required inputs.
 For each \( v \) with parent connections \( e = (u, v) \), execution on \( M \) follows:   
\begin{align}
  s_M(v) &= 
  \begin{cases}
    s_M(u) + C_M(u), \\
    \quad \text{if } x_M(v) - child(x_M(u)) = \emptyset, \\
    \max \big( s_R(u) + C_R(u) + T_{\text{RM}}(u, v), \\
    \quad\quad s_M(u) + C_M(u) \big), \\
    \quad \text{if } x_M(v) - child(x_M(u)) \neq \emptyset.
  \end{cases}
\end{align}
Here, \( \text{child}(x(u)) \) denotes the set of operations consuming outputs of \( x(u) \) according to their data dependencies, and the transmission volume in \( T_{\text{MR}}(u, v) \) is given by \( x_M(v) - \text{child}(x_M(u)) \). 
When \( x_M(v) - \text{child}(x_M(u)) = \emptyset \), all required inputs for \( x_M(v) \) are produced by operations on \( M \); otherwise, \( x_M(v) \) also depends on outputs from operations executed on \( R \).
Similar constraints apply to \( R \).



 



\myparagraph{Computational Efficiency}
To mitigate excessive kernel launch overhead, stemming from input fragmentation when \xxxschedule{} processes minimal units, and maximize GPU server resource utilization for faster overall inference, \xxx{} employs operation batching, instead of asynchronous execution~\cite{bin2024coacto}. 
This technique consolidates multiple fine-grained operations into a single system-level kernel launch, thereby preventing redundant invocations.
By managing the partitioning and reconstruction of input/output tensors for each operator, this approach preserves fine-grained control over local operations while minimizing redundant executions.
To formalize this, for all operators \( v \in V \): $x_M(v) = \left\{ i \mid i \in [\min(x_M(v)), \max(x_M(v))] \right\}$, $x_R(v) = \left\{ i \mid i \in [\min(x_R(v)), \max(x_R(v))] \right\}$.



\myparagraph{Transmission Efficiency}
Substantial transmission costs in Mobile Edge Computing (MEC) necessitate the minimization of redundant data transfers. 
Leveraging the insight from layer partitioning methods~\cite{hu2019dynamic}, whereby certain intermediate DNN layers yield significantly smaller outputs than raw input data, we impose the following constraint to curtail this overhead.
This also substantially reduces the search space, significantly accelerating the solving process of \xxxschedule{}.  
For layers \( u \) whose outputs exceed the raw input size (\( u \in \Pi \)), with all \( e = (u, v) \in E \): $x_M(v) - \text{child}(x_M(u)) = \emptyset$, and $x_R(v) - \text{child}(x_R(u)) = \emptyset$.

\subsubsection{Solution Algorithm}



As the search space increases from \( O(n) \) (layers) to \( O(n^2) \) (operations), \xxxschedule{} employs differential evolution~\cite{qin2008differential}, leveraging its robust global search capability and parallelism to efficiently solve the nonlinear, non-convex scheduling problem.
For a given hardware configuration, bandwidth emerges as the primary determinant for partitioning local operations. 
This insight allows \xxxschedule{} to significantly constrain the search space, thereby circumventing the substantial exploration overhead characteristic of methods like reinforcement learning~\cite{ji2022trajectory}.
Consequently, optimized scheduling plans only require recalculation when the underlying hardware changes (e.g., an upgraded mobile accelerator or GPU server). 
While generating the scheduling plans for a model takes a few minutes, this computation is performed offline within the Scheduling Optimization stage, imposing no impact on runtime inference performance.

\subsection{Adaptive Control Mechanism}
\label{sec:adaptive}
The adaptive control mechanism of \xxx{} operates on both the mobile device client and the GPU server (Fig.~\ref{fig:architecture}, \textbf{\textcircled{5}}). 
A backend network monitor on the mobile device ascertains TCP bandwidth at the transport layer; its energy consumption is negligible compared to that of inference computation. 
This mechanism enforces a stateless execution pattern: the mobile device and the GPU server synchronize current bandwidth and select an identical scheduling plan prior to each inference. 
The core assumption of \xxxschedule{} is stable bandwidth throughout an individual inference request, which obviates the need to switch scheduling plans for ongoing inference.
Leveraging model replicas on the GPU server alongside pre-computed scheduling plans, the adaptive control within \xxx{} facilitates seamless, delay-free transitions between different schedules, enabling instant adaptation to network variations. 
Consequently, \xxx{} maintains robust performance even as mobile devices transition between networks, since its scheduling decisions depend solely on the transport-layer bandwidth between the mobile device and the GPU server.

The detailed algorithms constituting this adaptive control mechanism are presented in Alg.~\ref{alg:client} and Alg.~\ref{alg:server}.
During runtime, \xxx{} estimates real-time bandwidth (line 1 in both algorithms) and selects a suitable scheduling plan. 
It then computes the designated local operations (line 5 in both algorithms), producing outputs that represent the computation results of these local operations or an empty set if no local computation is performed.
Leveraging operation batching from \xxxschedule{}'s computational efficiency constraints, \xxx{} merges all operations within the same operator ($x^b_M(u)$) into a single kernel launch. 
This launch executes once the required inputs are ready, prioritizing inference correctness over strict adherence to the scheduled execution time ($s_M(u)$). 
The associated blocking overhead, primarily from tensor combining, is optimized in \xxxschedule{}.

\begin{algorithm}[!t]
\caption{\xxx{} client at runtime stage}
\label{alg:client}
\begin{algorithmic}[1]
\small
\Statex \textbf{Input:} Data input for inference $\texttt{input}$; DNN model $\texttt{model}$
\Statex \textbf{Output:} The inference result $\texttt{ret}$ 
\Statex \textbf{Parameter:} $\texttt{Input(i)}$,$\texttt{Output(i)}$: input and output of layer $i$; $x^b_M$, $x^b_R$: schedule plan under the $b$ bandwidth. 
\State  $b \gets$ \Call{TestBandwidth}{\texttt{}}
\State  $\texttt{Input(0)} \gets \texttt{input}$
\ForAll {layer $u$ in model}
\If {$Input(u) \neq \emptyset$ and $x^b_M(u) \neq \emptyset $}
\State $\texttt{output(u)} \gets$ \Call{compute}{$\texttt{Input(u)},x^b_M(u)$}
\EndIf
\ForAll {edge $e = (u,v) \in E$}
\If {$(x^b_M(v) - \texttt{child}(x^b_M(u)) \neq \emptyset$}
\State $\texttt{Input(v)} \gets$  \Call{combine}{$\texttt{Output(u)},\texttt{Receive()}$}
\EndIf
\If {$x^b_R(v) - \texttt{child}(x^b_R(u)) \neq \emptyset$}
\State \Call{Send}{$\texttt{Output(u)}, x^b_R(v) -  \texttt{child}(x^b_R(u))$}
\EndIf
\EndFor
\EndFor
\State \Return $ret \gets \texttt{Output(t)}$
\end{algorithmic}
\end{algorithm}

\begin{algorithm}[!t]
\caption{\xxx{} server at runtime stage}
\label{alg:server}
\begin{algorithmic}[1]
\small
\State  $b \gets$ \Call{Receive}{\texttt{}}
\State  $\texttt{Input(0)} \gets \emptyset$
\ForAll {layer $u$ in model}
\If {$Input(u) \neq \emptyset$ and $x^b_R(u) \neq \emptyset $}
\State $\texttt{output(u)} \gets$ \Call{compute}{$\texttt{Input(u)},x^b_R(u)$}
\EndIf
\ForAll {edge $e = (u,v) \in E$}
\If {$x^b_M(v) - \texttt{child}(x^b_M(u)) \neq \emptyset$}
\State \Call{Send}{$\texttt{Output(u)}, x^b_M(v) -  \texttt{child}(x^b_M(u))$}
\EndIf
\If {$(x^b_R(v) - \texttt{child}(x^b_R(u)) \neq \emptyset$}
\State $\texttt{Input(v)} \gets$  \Call{combine}{$\texttt{Output(u)},\texttt{Receive()}$}
\EndIf
\EndFor
\EndFor
\end{algorithmic}
\end{algorithm}

If the scheduling plan requires data transmission, the server sends the computed outputs to the client (line 9 in Alg.\ref{alg:server}), which integrates them with its results for subsequent operators (line 9 in Alg.\ref{alg:client}). Similarly, if data needs to be sent from client to server, the client transmits the outputs (line 12 in Alg.\ref{alg:client}), and the server integrates them accordingly (line 12 in Alg.\ref{alg:server}). Otherwise, the outputs of the current operator directly serve as inputs for subsequent operators. Once all operations are completed, the final inference result remains on the client, ready for upper-layer applications (line 16 in Alg.\ref{alg:client}).

\section{Implementation}
\label{sec:implement}

In this section, we first elaborate on the implementation of \xxx, and we then introduce the experiment setup.

\subsection{System Implementation}
\subsubsection{Software}
We implemented \xxx{} using Python and PyTorch~\cite{pytorch}, integrating it seamlessly into existing deep learning workflows. 
The implementation hooks into the model’s forward method, where the first forward pass profiles the model using PyTorch’s default profiler and execution schedule. 
Based on this profiling data, \xxx{} then intercepts and parallelizes all subsequent forward calls according to the optimized execution plan. 
With a lightweight integration requiring only three lines of code, \xxx{} ensures ease of deployment in existing applications.

\subsubsection{Hardware Testbed}
\begin{figure}[!t]
\vspace{-10pt}
\setlength\abovecaptionskip{6pt}
\setlength\belowcaptionskip{0pt}
    \centering
    \subfloat[Four-wheeled robot\label{fig:robot}]{\includegraphics[width=0.45\linewidth]{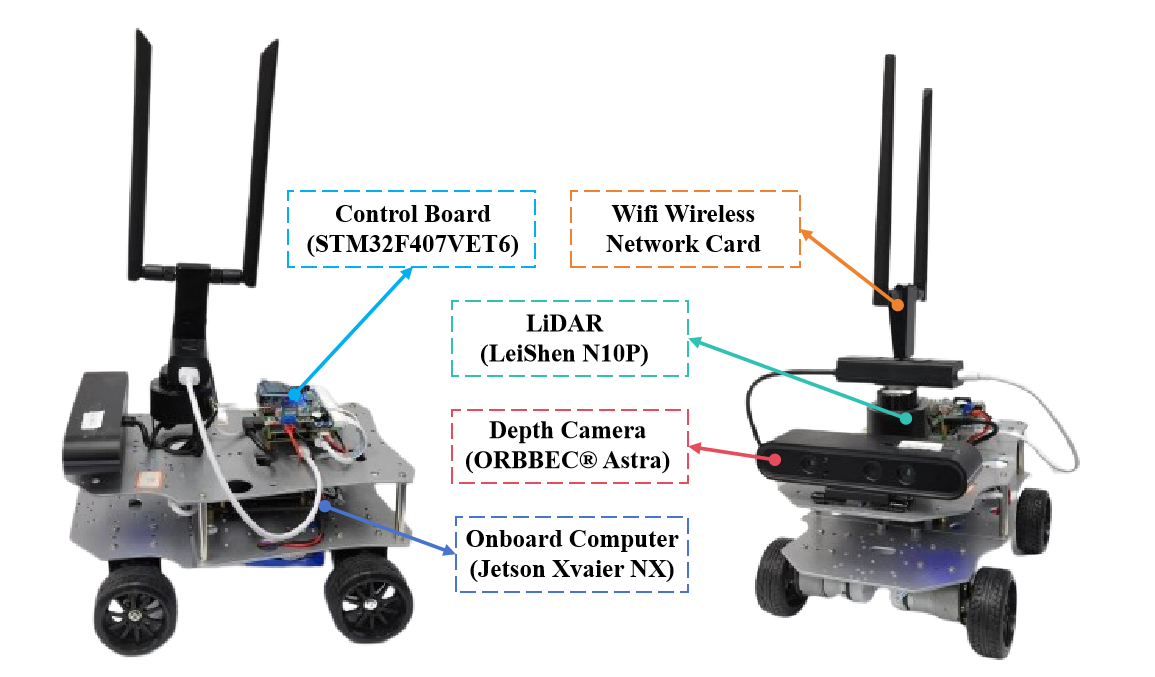}}
    \hfil
    \subfloat[Air-ground robot\label{fig:airground}]{\includegraphics[width=0.45\linewidth]{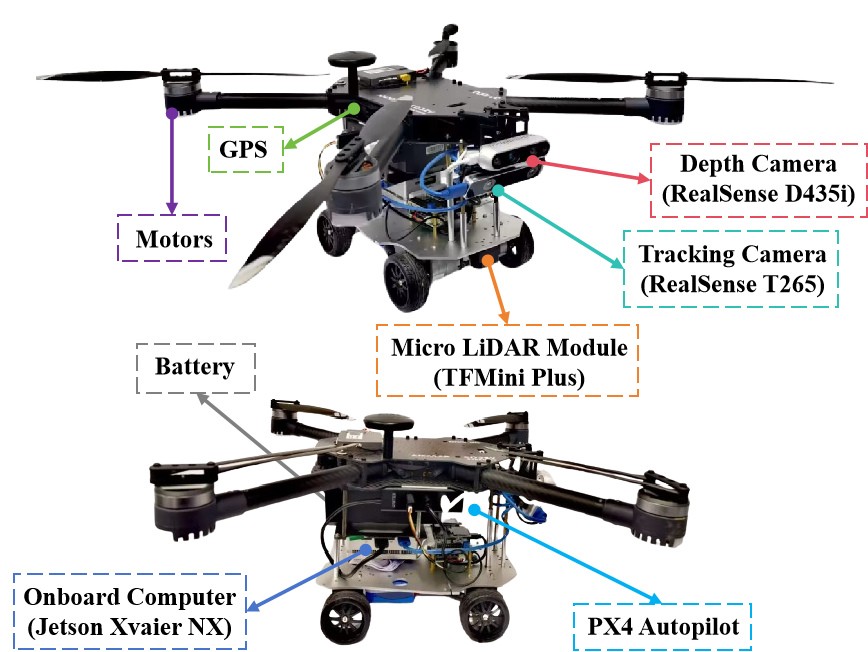}}
    \caption{\small The detailed composition of the robot platforms.}
    \label{fig:robots}
\end{figure}

We evaluated \xxx{} on two customized robotic platforms: a four-wheeled ground robot (Fig.~\ref{fig:robot}) and an air-ground hybrid robot (Fig.~\ref{fig:airground}). 
Each robot is equipped with a Jetson Xavier NX (8GB)~\cite{jetsonnx}, enabling on-board model inference with local computation. 
The system runs Ubuntu 20.04 with ROS Noetic and uses a dual-band USB network adapter (MediaTek MT76x2U) for wireless communication. 
Detailed hardware and sensor configurations are shown in Fig.~\ref{fig:robots}. 
The GPU server is a PC equipped with an Intel i7-7700K CPU and an NVIDIA GeForce RTX 3080 GPU. 
It connects to the robots via Wi-Fi 6 on an 80 MHz channel at 5 GHz.  

Tab.~\ref{tab:energydefault} summarizes the robots' on-board energy consumption (excluding motor power) in different states: inference (full GPU utilization, including CPU/GPU power), communication (communication with the GPU server, including wireless network card energy consumption), standby (no tasks to execute). 
Each Jetson Xavier NX is powered by a 21.6 Wh battery, sustaining up to 1.6 hours of continuous model inference. 
We continuously log the robot's instantaneous on-board power draw (in Watts) at 1-second intervals, utilizing the back-end power consumption and performance monitoring methodology from~\cite{jetsonnx}.
Subsequently, the energy consumption per inference (in Joules) is precisely calculated by integrating this power draw profile over the exact duration of each inference, determined by its start and end timestamps.

\begin{table}[!t]
\setlength\abovecaptionskip{6pt}
\setlength\belowcaptionskip{-2pt}
\centering
\begin{tabular}{|c|c|c|c|}
\hline
        & inference & communication & standby \\ \hline
   Energy (Watt) &     13.35        &       4.25        &    4.04   \\ \hline
\end{tabular}
\caption{\small Power draw (Watt) of our robot in different states.}
\label{tab:energydefault}
\end{table}

\subsection{Experiment Setup}

\begin{figure}[!t]
\setlength\abovecaptionskip{6pt}
\setlength\belowcaptionskip{-2pt}
    \centering
    \includegraphics[width=0.9\linewidth]{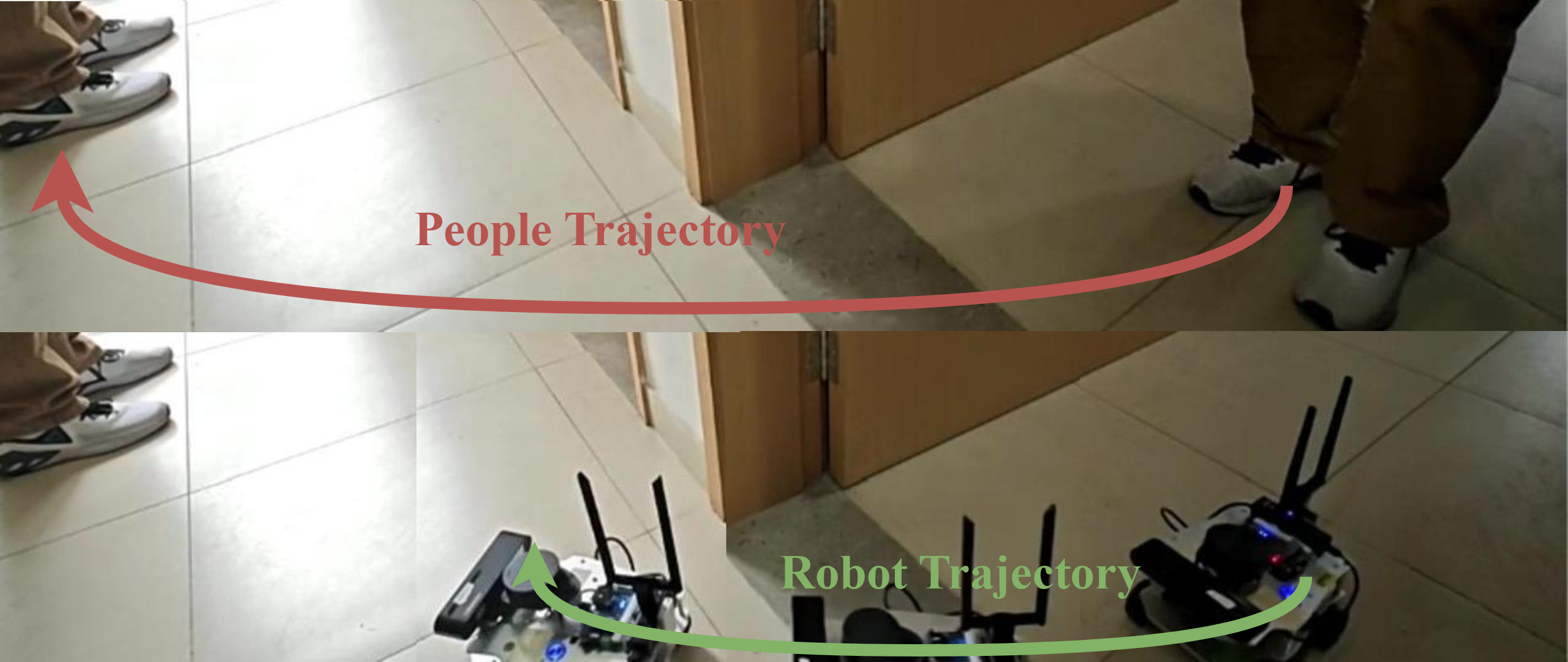}
    \caption{\small Kapao~\cite{kapao}, a real-time people-tracking application on our four-wheeled robot with a CNN-based human keypoint detection model.}
    \label{fig:kapao}
\end{figure}

\begin{figure}[!t]
\setlength\abovecaptionskip{6pt}
\setlength\belowcaptionskip{-2pt}
    \centering
    \includegraphics[width=0.9\linewidth]{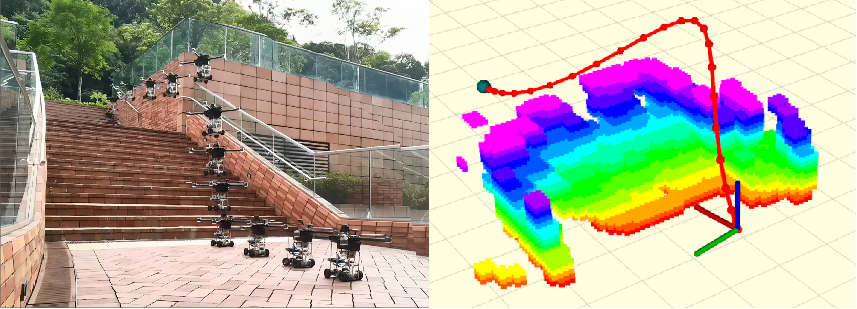}
    \caption{\small AGRNav~\cite{agrnav}, an autonomous navigation application on our air-ground robot with a CNN-based 3D semantic scene completion model.}
    \label{fig:agrnav}
\end{figure}

\subsubsection{Task}
We evaluated two typical real-world robotic applications on our testbed: Kapao~\cite{kapao} (Fig~\ref{fig:kapao}) and AGRNav~\cite{agrnav} (Fig~\ref{fig:agrnav}). 
We also evaluated several models common to mobile devices with their implementation from Torchvision~\cite{torchvision} on a larger scale to further corroborate our observations and findings: VGGNet~\cite{simonyan2015deep}, ConvNeXt~\cite{woo2023convnext}, ResNet~\cite{targ2016resnet}, DenseNet~\cite{huang2018densely}.

\subsubsection{Emulation Environments}
We evaluated two real-world environments: indoors (robots move in our laboratory with desks and separators interfering with wireless signals) and outdoors (robots move in our campus garden with trees and bushes interfering with wireless signals, resulting in lower bandwidth). 
The corresponding bandwidths between the robot and the GPU server in indoors and outdoors scenarios are shown in Fig.~\ref{fig:bandwidth}.

\subsubsection{Baselines}
To comprehensively evaluate the performance of \xxx, we conducted comparative experiments against several baseline approaches:

\begin{itemize}
    \item \textbf{Device-only inference (``Device-only'')}: A conventional setup where the entire model is deployed and executed on the mobile device.  
    
    \item \textbf{Server-only inference (``Server-only'')}: A cloud-based approach in which the full model runs on a GPU server. While both Device-only and Server-only represent extreme cases of layer partitioning, we evaluate them separately to highlight the benefits of other baselines in terms of inference latency and energy efficiency.  
    
    \item \textbf{DSCCS}~\cite{liang2023dnn}: A state-of-the-art layer partitioning method designed for inference acceleration. To ensure a fair comparison and eliminate the influence of parameter variations, we enhance DSCCS with \xxx’s adaptive control mechanism, allowing it to dynamically adjust to wireless network fluctuations.  
    
    \item \textbf{SPSO-GA}~\cite{chen2021energy}: An advanced layer partitioning method that optimizes energy consumption while satisfying timing constraints. We configure SPSO-GA with a 1 Hz deadline constraint, corresponding to the minimum frequency required for effective robotic motion control. As with DSCCS, we integrate \xxx’s adaptive control mechanism into SPSO-GA to allow real-time adaptation to network variations.  
\end{itemize}  

For consistency and accurate performance comparison, all systems are implemented without model compression, transmitting raw model activations to maintain accuracy.  
The Profiling and Scheduling Optimization stages of \xxx{} and all baselines run offline in advance.

\section{Performance Evaluation}
\label{sec:evaluation}

In this section, we evaluate the performance of \xxx{} from three research aspects: \begin{inparaenum}[i)] \item comparisons with baseline systems to demonstrate the superiority of \xxx{} in inference time and energy consumption for real-world mobile applications; \item investigating the \xxxschedule{} of \xxx{} through micro-event analysis; \item sensitivity studies to analyze system performance under varying network bandwidth variations, model structures, and equipment computational powers\end{inparaenum}.

\begin{figure}[!t]
\setlength\abovecaptionskip{6pt}
\setlength\belowcaptionskip{0pt}
    \centering
    \includegraphics[width=0.98\linewidth]{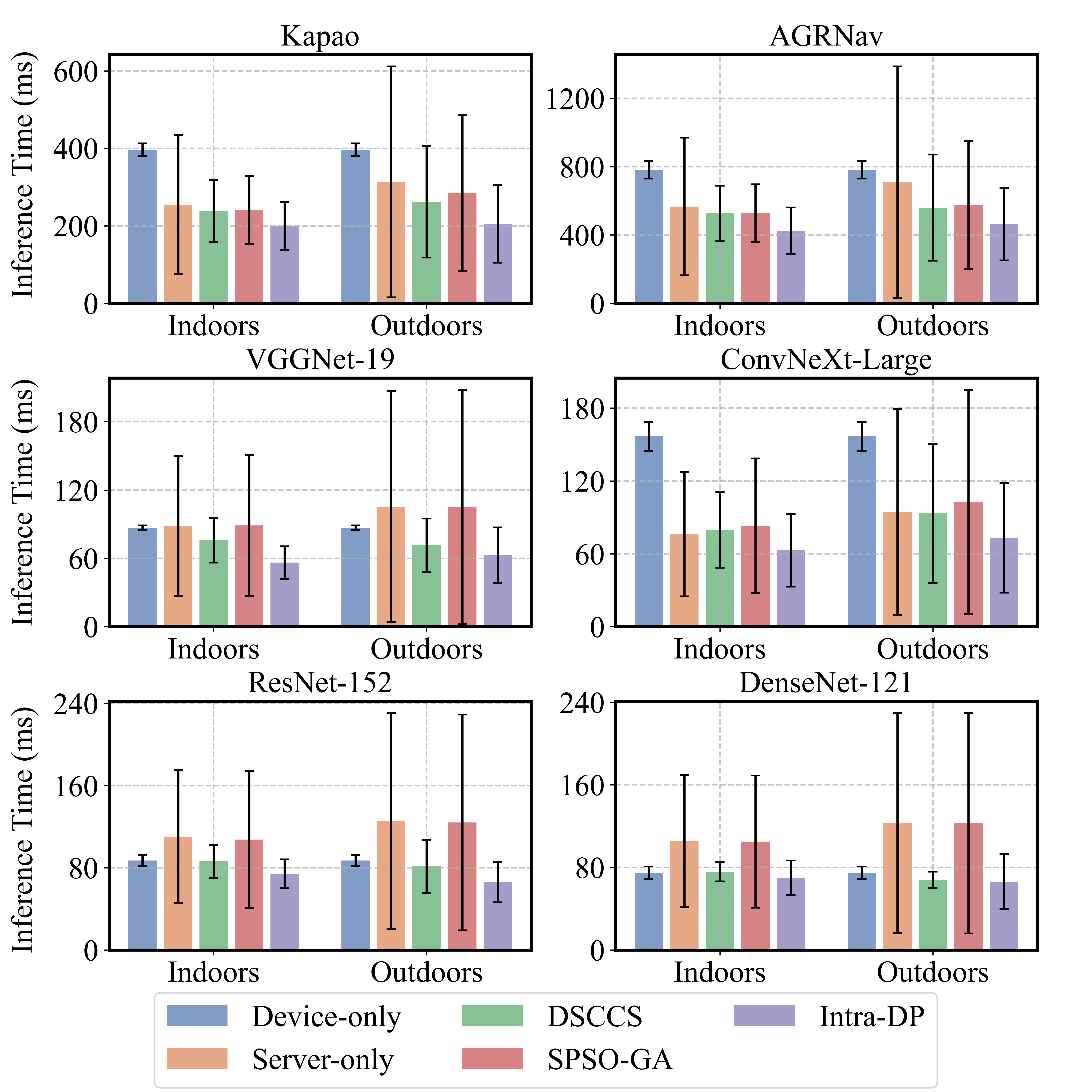}
    \caption{\small Inference time for different models across various environments and systems.}
    \label{fig:end2end_inference}
\end{figure}

\subsection{Superiority of \xxx}
\subsubsection{Inference Time}
Fig.~\ref{fig:end2end_inference} demonstrates the superior performance of \xxx{} to four baseline methods across various tasks and environments, achieving significant speedup in inference time (up to 50\%  for indoors and up to 48\% for outdoors). 
\xxx{} outperforms the closest baseline DSCCS by 8\% to 26\% indoors and 8\% to 22\% outdoors. 
These improvements primarily result from \xxxparallel{}, which enables parallel execution and overlaps computation with transmission across local operations within the same inference request, and \xxxschedule{}, which optimizes scheduling under varying network conditions. 
Notably, all offloading-based methods, including server-only, two layer partitioning methods, and \xxx{}, exhibit greater standard deviations and longer inference times outdoors due to severe bandwidth fluctuations, as shown in Fig.~\ref{fig:bandwidth}. 
Although the performance gain is less pronounced on DenseNet, this is analyzed in our sensitivity study in Sec.~\ref{sec:sensitivity_structure}. 
Additionally, a micro-event analysis in Sec.~\ref{sec:microevent} further explains the efficiency of \xxx{} in inference.


\subsubsection{Energy Consumption}

\begin{figure}[!t]
\setlength\abovecaptionskip{6pt}
\setlength\belowcaptionskip{0pt}
    \centering
    \includegraphics[width=0.98\linewidth]{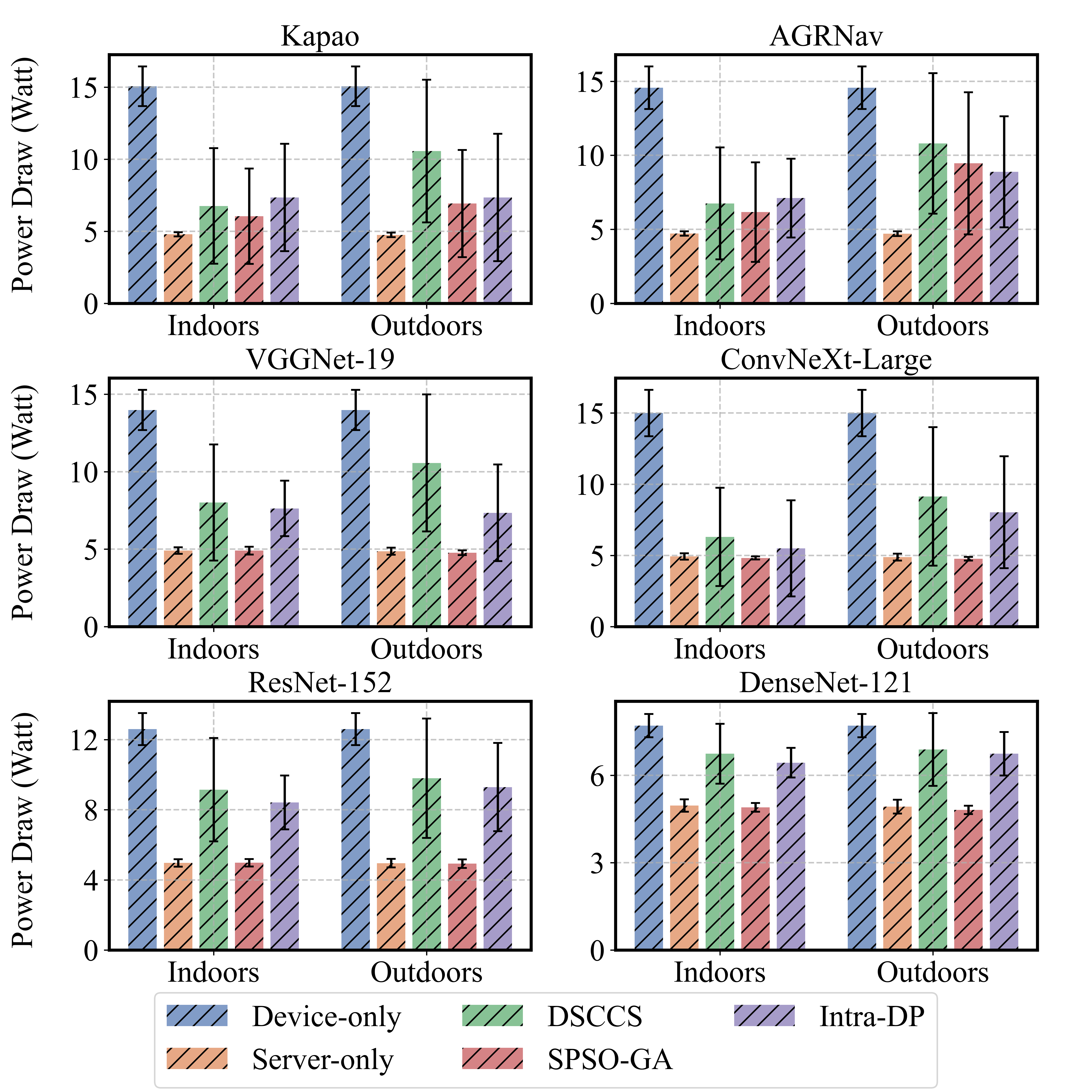}
    \caption{\small Power draw for different models across various environments and systems.}
    \label{fig:end2end_energy_per_unit_time}
\end{figure}

Fig.~\ref{fig:end2end_energy_per_unit_time} presents the power draw across different methods and scenarios. 
The device-only approach consumes the most energy due to the intensive computational load on robots, while the server-only approach achieves the lowest energy consumption by offloading all computations to the GPU server.
Among the partial-offloading methods, SPSO-GA demonstrates better energy efficiency than DSCCS due to its energy-oriented optimization, whereas \xxx{} exhibits slightly higher energy consumption due to occasional re-computation of local operations. 
The power draw between \xxx{} and DSCCS remains nearly the same, indicating that the computation-communication overlap in \xxx{} does not introduce significant additional energy costs. Tab.~\ref{tab:energydefault} further supports this conclusion, showing that robots continue consuming 95\% of their active power even when idle, mainly due to static energy consumption in components such as the CPU, GPU, and memory~\cite{kim2003leakage}. In contrast, the wireless network card consumes only 0.21 W during transmission compared to 13.35 W during computation.


\begin{figure}[!t]

\setlength\abovecaptionskip{6pt}
\setlength\belowcaptionskip{0pt}
    \centering
    \includegraphics[width=0.98\linewidth]{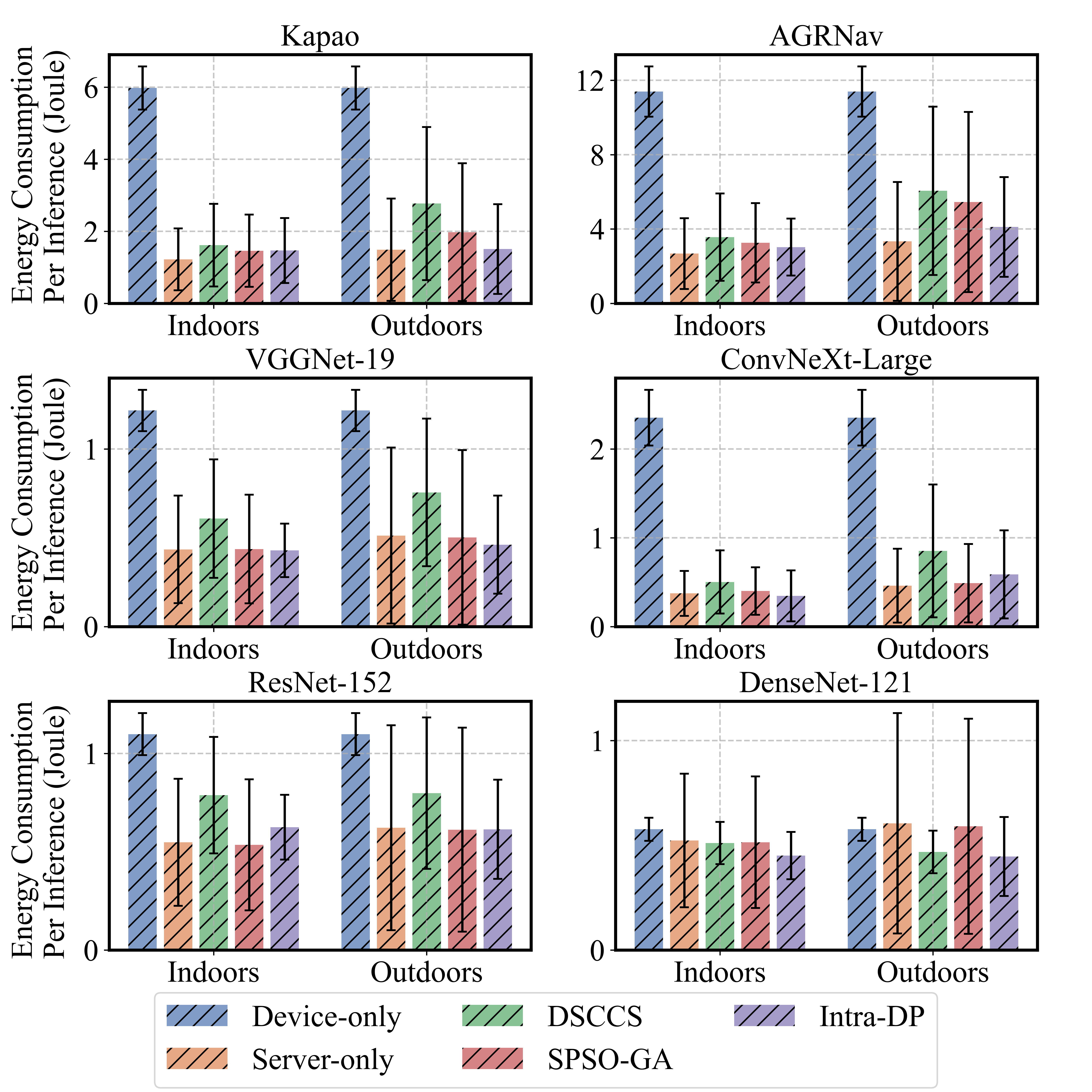}
    \caption{\small Energy consumption per inference request for different models across various environments and systems.}
    \label{fig:end2end_energy_per_inference}
\end{figure}

Fig.~\ref{fig:end2end_energy_per_inference} presents the energy consumption per inference request across different methods and scenarios. 
Although \xxx{} exhibits relatively higher power draw in Fig.~\ref{fig:end2end_energy_per_unit_time}, it achieves significantly lower energy consumption per inference request across all models and environments, reducing energy usage by up to 75\% indoors and up to 72\% outdoors due to its fast inference time. 
Notably, compared to the most energy-efficient server-only approach, \xxx{} increases energy consumption per inference request by at most 20\% while reducing inference time by up to 42\% and greatly reduced long tail delays.

\subsection{Micro-Event Analysis}
\label{sec:microevent}

 \begin{figure}[t!]
 \vspace{-5pt}
\setlength\abovecaptionskip{-20pt}
\setlength\belowcaptionskip{0pt}
\centering
\includegraphics[width=0.98\linewidth]{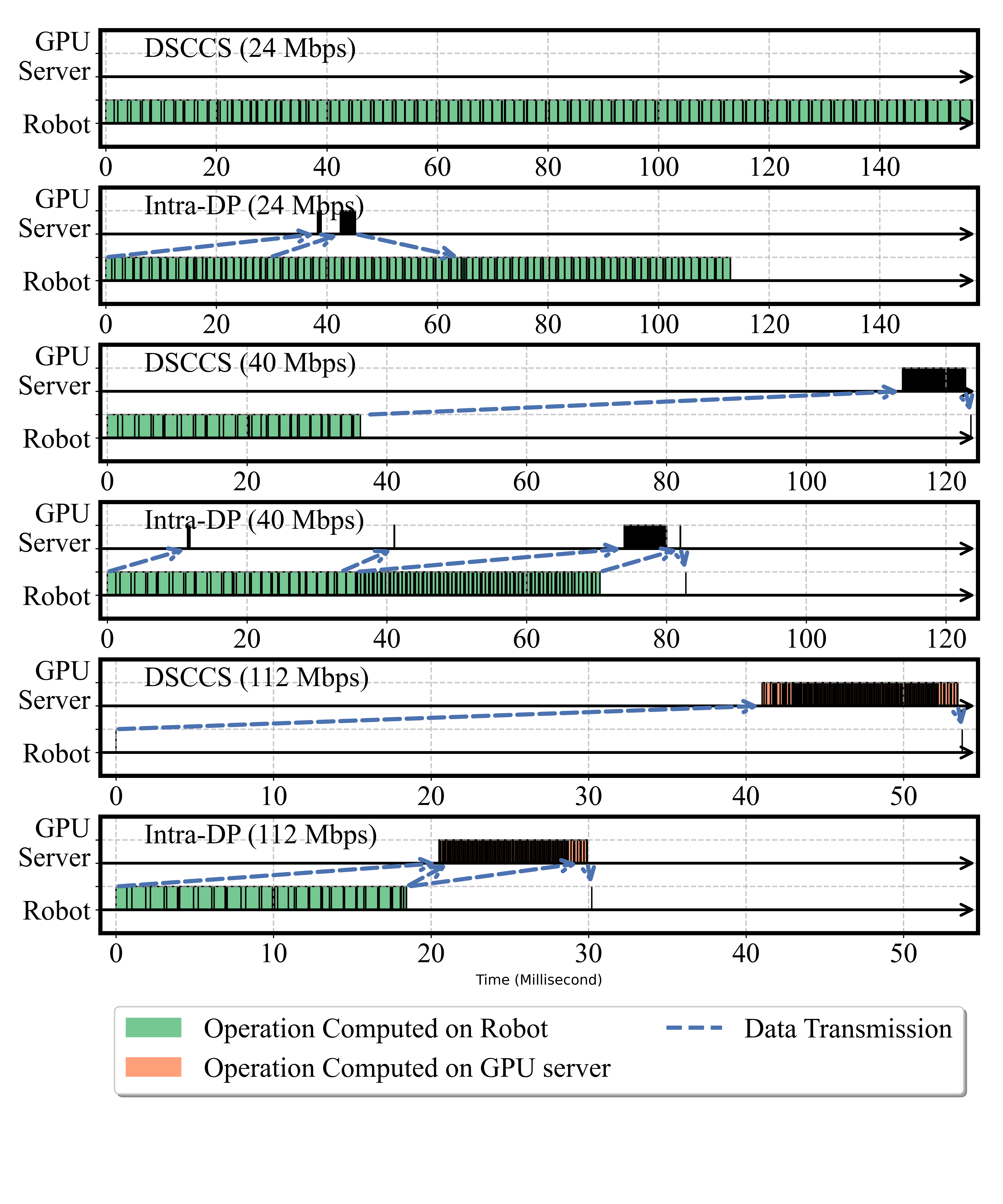}
\caption{\small Snapshots of schedule plan of \xxx{} and baseline during runtime under various network bandwidth.}
\label{fig:microevent}
\end{figure}

\begin{figure}[!t]
\setlength\abovecaptionskip{-5pt}
\setlength\belowcaptionskip{0pt}
    \centering
    \includegraphics[width=0.98\linewidth]{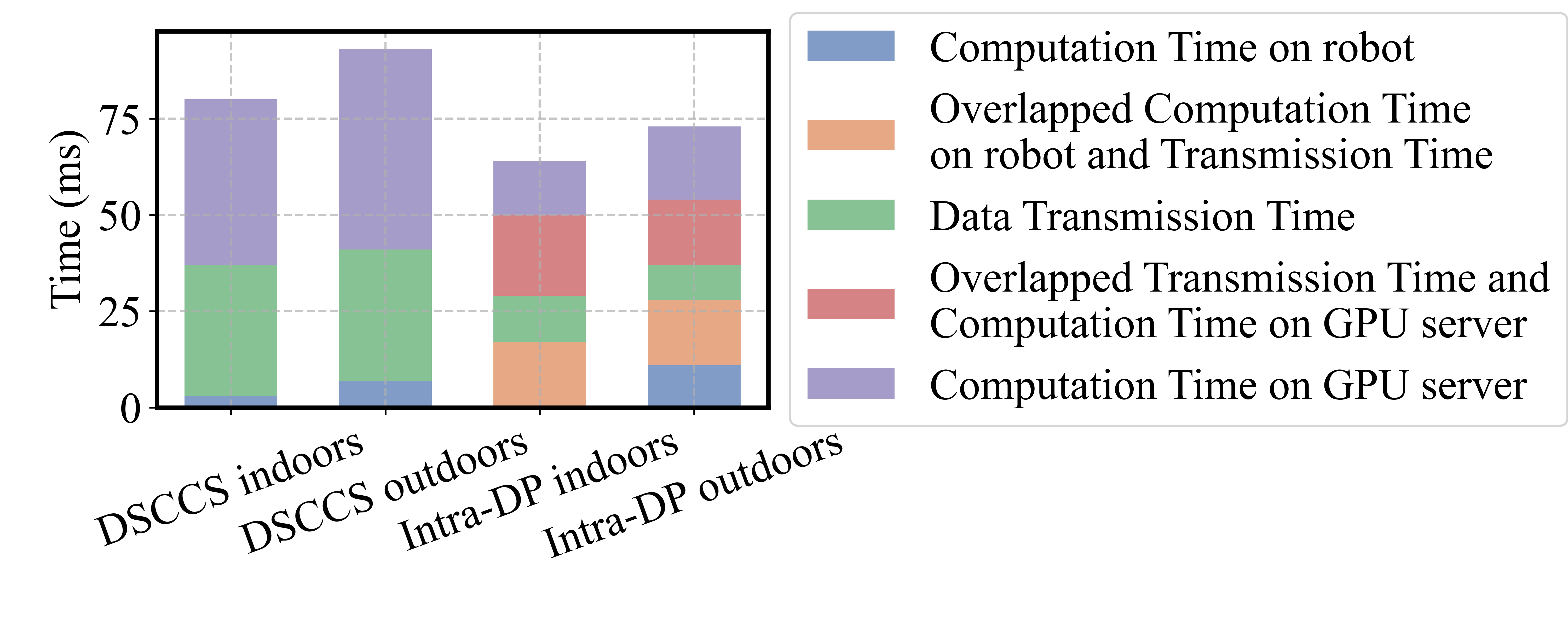}
    \caption{\small Breakdown of each phase of the inference process.}
    \label{fig:breakdown}
\end{figure}

Next, we present a detailed investigation of \xxx{}'s superior inference speed through micro-event analysis, focusing on the effectiveness of \xxxschedule{} compared to baseline methods.
\subsubsection{Detailed Schedule Plan}
To quantify the performance advantages of \xxx{}, we conduct a comparative analysis with DSCCS by recording real-time scheduling plans for ConvNeXt under varying bandwidth conditions, as shown in Fig.~\ref{fig:microevent}. 
The primary performance gain of \xxx{} stems from the parallel execution of local operations, enabling concurrent computation on the robot, data transmission, and computation on the GPU server.  
With the integration of \xxx{}'s adaptive control mechanism, DSCCS can dynamically select the optimal layer partitioning plan based on real-time network bandwidth.  
Under low-bandwidth conditions, DSCCS shifts more computation to the device, trading increased energy consumption for reduced tail latency. 
While DSCCS degrades to device-only, \xxx{} maintains its distributed execution capability by leveraging an expanded parallel scheduling space enabled by \xxxparallel{}, requiring less bandwidth for GPU server acceleration. 
Conversely, in high-bandwidth scenarios, DSCCS transitions to server-only, whereas \xxx{} leverages \xxxschedule{} to optimize network and GPU server utilization for faster inference.

\subsubsection{Breakdown}
We also provide a detailed breakdown of each phase of the inference process throughout the ConvNeXt experiment, as illustrated in Fig.~\ref{fig:breakdown}. 
In DSCCS, the transmission time accounts for up to 44.2\% of the total inference time, underscoring the significant transmission bottlenecks encountered by existing layer partitioning methods, even when employing state-of-the-art strategies.  
Unlike layer partitioning methods, which execute each phase sequentially, \xxx{} enables the transmission phase, the robot computation phase, and the GPU server computation phase to run in parallel, resulting in faster inference times.
While \xxx{} incurs additional communication overhead from its fine-grained transmission of local operators, it effectively shortens overall completion time. 
This is achieved by overlapping computation with communication and initiating data transfer early (Fig.~\ref{fig:microevent}), leading to superior performance over traditional layer partitioning methods.

\subsection{Sensitivity Studies}
\label{sec:sensitivity}

\subsubsection{Network Bandwidth}
\label{sec:sensitivity_bandwidth}
\begin{figure}[!t]
\setlength\abovecaptionskip{-20pt}
\setlength\belowcaptionskip{0pt}
    \centering
    \includegraphics[width=0.98\linewidth]{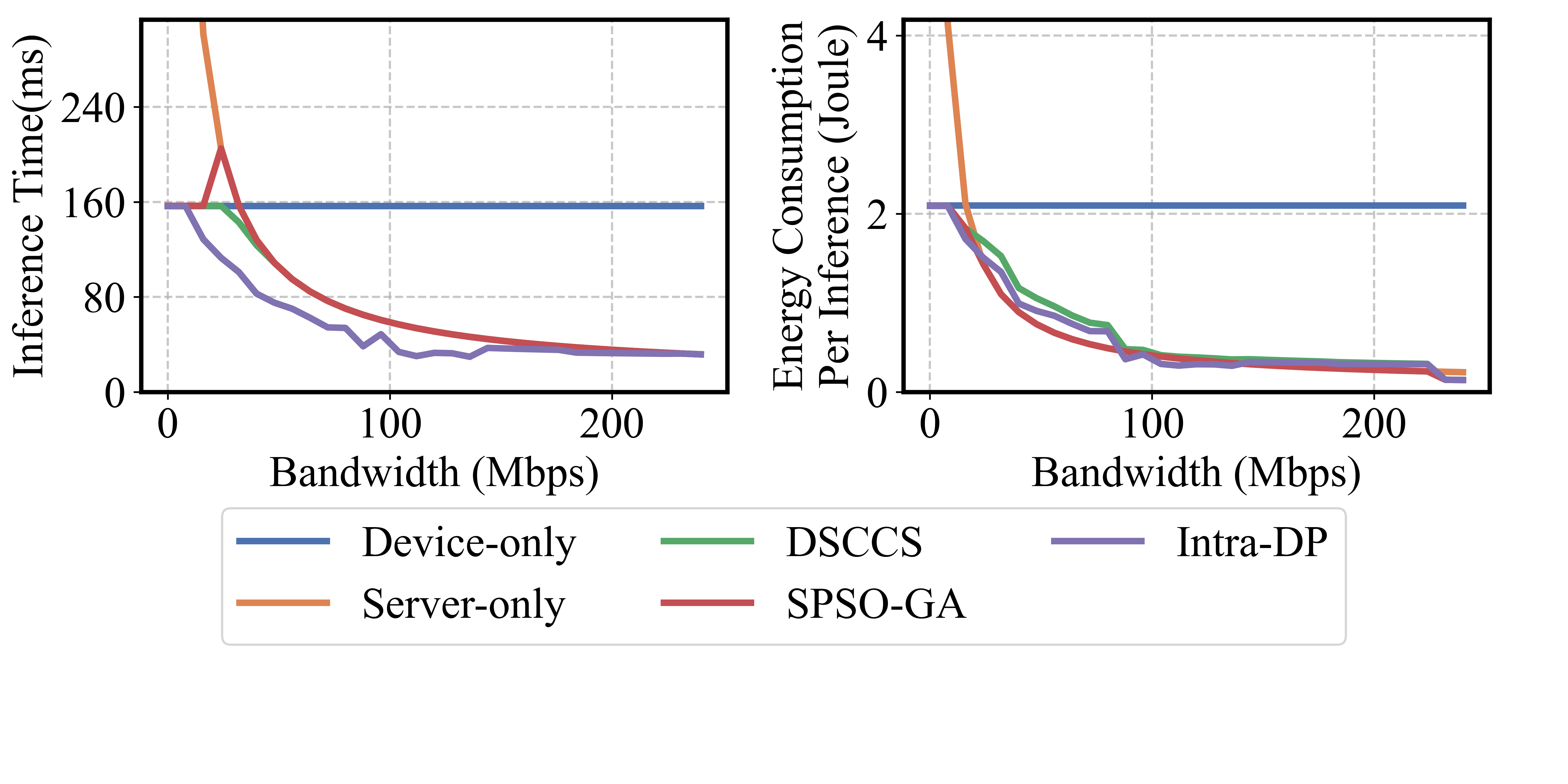}
    \caption{\small Performance comparison of \xxx{} and baselines under different network bandwidth conditions.}
    \label{fig:sensitivity_bandwidth}
\end{figure}

To systematically evaluate \xxx{}'s performance under varying bandwidth conditions in MEC, we leveraged the Linux Traffic Control (tc) tool~\cite{beshay2015fidelity} over wired Ethernet to create controlled test scenarios, enabling comprehensive comparisons with baseline methods.  
As shown in Fig.~\ref{fig:sensitivity_bandwidth}, \xxx{} maintains superior performance across a wider range of network conditions, thanks to the expanded scheduling space enabled by \xxxparallel{}.  
When the bandwidth is extremely low (near 0 Mbps), \xxx{} also degrades to device-only; however, its threshold bandwidth for degradation is significantly lower than that of DSCCS. 
Similarly, under extremely high bandwidth conditions, \xxx{} degrades to server-only, but its degradation threshold remains much higher than DSCCS. 
This highlights \xxx{}'s extensive parallel optimization space, enabling more efficient execution across diverse network conditions.  
Notably, the bandwidth thresholds at which layer partitioning methods and \xxx{} transition between device-only and server-only depend on model architecture and device computational power. 
Furthermore, the performance gain of \xxx{} in Fig.~\ref{fig:end2end_inference} and Fig.~\ref{fig:end2end_energy_per_inference} is smaller than in Fig.~\ref{fig:sensitivity_bandwidth} due to inaccurate bandwidth estimation during runtime.

\subsubsection{Model structure}
\label{sec:sensitivity_structure}
\begin{figure}[!t]
\setlength\abovecaptionskip{-5pt}
\setlength\belowcaptionskip{0pt}
    \centering
    \includegraphics[width=0.98\linewidth]{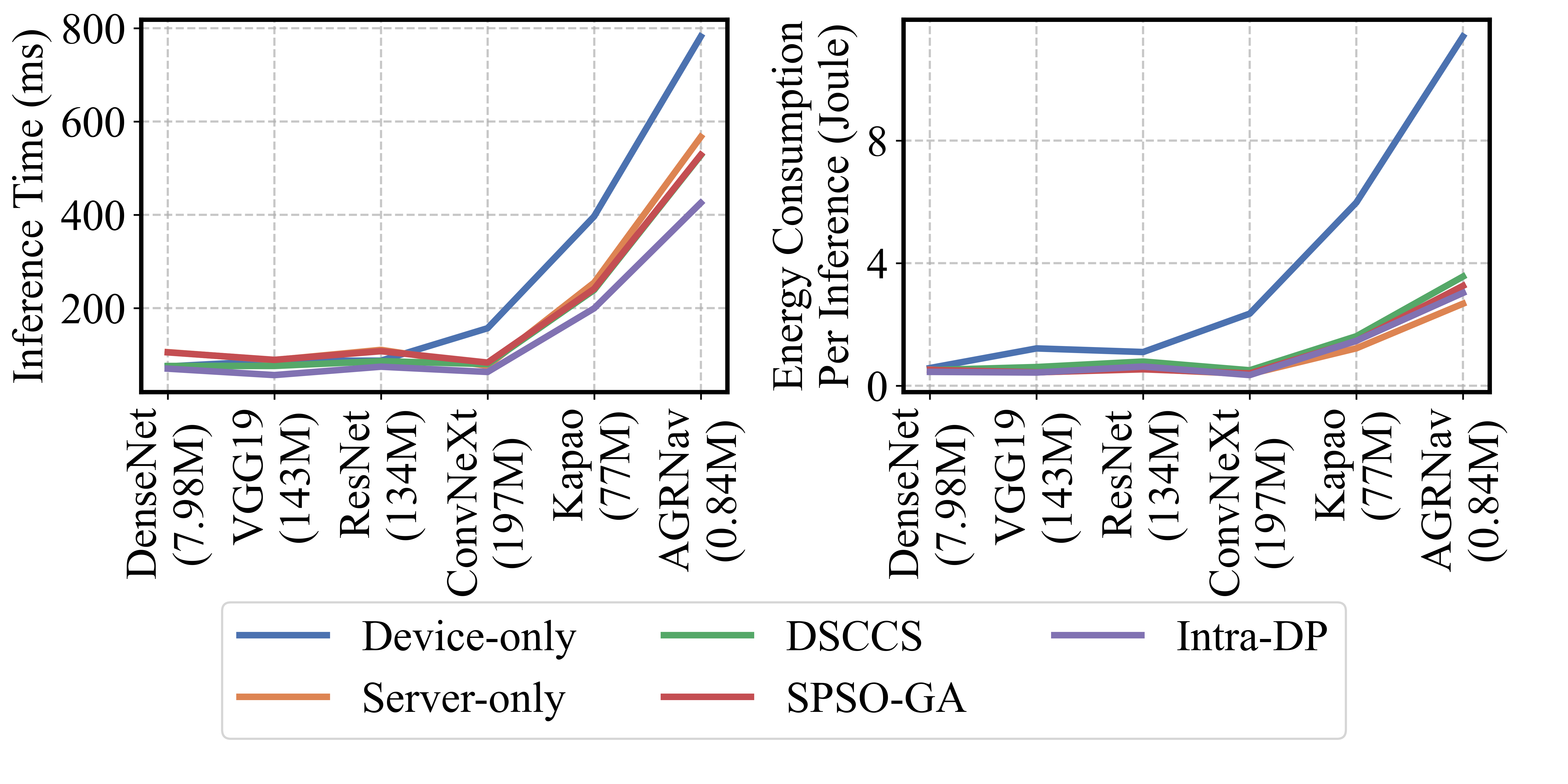}
    \caption{\small Performance comparison of \xxx{} and baselines with varying model structures.}
    \label{fig:sensitivity_model}
\end{figure}

As shown in Fig.~\ref{fig:sensitivity_model}, offloading methods (except device-only) achieve greater performance gains for models with higher computational demands but perform worse than device-only on DenseNet.  
This indicates that offloading is more beneficial for computationally intensive models, while models with lower computational demands may suffer from communication overhead, especially in bandwidth-constrained MEC systems.
While \xxx{} consistently outperforms other baselines, its performance gains over them are relatively smaller for models with fewer parameters.  
This observation can be attributed to the fact that \xxx{}'s performance improvements are primarily achieved through the parallel execution of local operations, making its effectiveness dependent on the model’s size and structure. 
When a model has fewer parameters, the number of local operations available for parallel execution is reduced, thereby limiting \xxx{}'s optimization potential for enhancing performance.

\begin{figure}[!t]
\setlength\abovecaptionskip{-30pt}
\setlength\belowcaptionskip{0pt}
    \centering
    \includegraphics[width=0.98\linewidth]{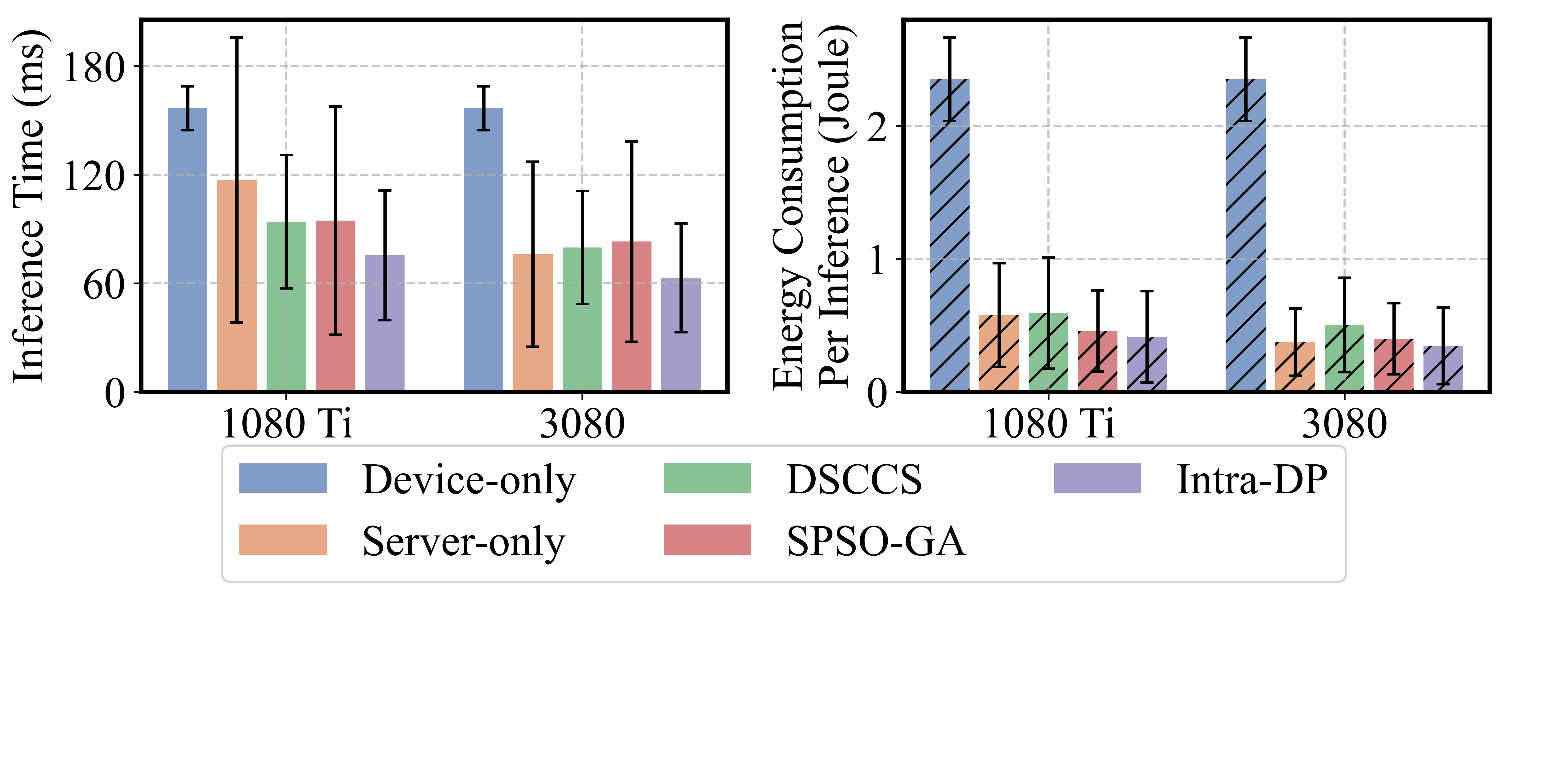}
    \caption{\small Performance comparison of \xxx{} and baselines on GPU servers with varying computational power.}
    \label{fig:sensitivity_computing_power}
\end{figure}

\subsubsection{Computational Power}
\label{sec:sensitivity_computing_power}
To assess \xxx{}'s performance under varying computational power disparities between the robot and a GPU server, we conducted evaluations using an NVIDIA GTX 1080 Ti. 
As illustrated in Fig.~\ref{fig:sensitivity_computing_power}, the performance advantage of offloading methods over device-only inference grows with increasing disparity. 
Notably, for any given computational gap, \xxx{} demonstrated the most significant improvement compared to other offloading approaches. This superior performance stems from its optimization of GPU server utilization via parallel execution of local operations. 
It is also pertinent to note that our experimental robot (Fig.~\ref{fig:device-inference}) possesses considerably higher computational power than typical mobile devices (e.g., smartphones), suggesting \xxx{}'s benefits could be even more pronounced on more resource-constrained platforms.

\section{Related work and discussion}
\label{sec:discussion}
\myparagraph{Limited bandwidth}
Due to hardware limitations, we evaluated \xxx{}'s performance only on commonly available Wi-Fi networks in robotic environments, rather than a broader set of wireless technologies (e.g., 5G). 
While different wireless networks vary in throughput and range due to distinct transmission protocols, they all suffer from weak connections and bandwidth fluctuations in practice~\cite{masiukiewicz2019throughput, ding2015performance, ren2018proportional}.
Under these conditions, \xxx{} proves beneficial. 
When GPU servers are deployed in commercial cloud environments, network congestion and routing inefficiencies further restrict available bandwidth~\cite{noormohammadpour2017datacenter}, making \xxx{} even more advantageous. 
Furthermore, as models scale and GPU computational power increases, \xxx{}'s advantage continues to grow, further highlighting its relevance.

\myparagraph{Inference Request Scheduling}
It schedules multiple DNN inference requests to optimize overall latency and energy consumption, leveraging existing collaborative inference methods for individual executions and employing various decision algorithms to determine optimal execution locations and timing. 
These decision algorithms are crucial for adapting to the unique transmission and computation demands inherent in diverse MEC scenarios. 
For instance, such approaches are applied in industrial IoT networks with replay selection~\cite{zhao2019novel}, space information networks with reconfigurable intelligent surfaces~\cite{cao2021converged}, distributed multi-node networks with integrated localization and computing of radio frequency signals~\cite{qi2024joint}, and heterogeneous vehicular networks with Vehicle-to-Everything communication technologies~\cite{xiong2021intelligent}. 
\xxx{} enhances this domain by introducing a higher-performance collaborative inference method tailored for these individual executions within MEC. 
Consequently, existing and future scheduling strategies can seamlessly integrate \xxx{} to capitalize on its improved efficiency for individual inference tasks, thereby boosting their overall effectiveness.

\myparagraph{Model Compression}
Quantization and model distillation are two most common model compression techniques for mobile devices. 
Quantization~\cite{gong2020vecq} reduces the precision of model weights and activations to lower computation costs, while model distillation~\cite{wang2021knowledge} trains a smaller model to mimic a larger one with fewer resources. 
Unlike these methods, which trade accuracy for efficiency, collaborative inference accelerates computation by intelligently distributing workloads without compromising accuracy.
Building on the significant performance gains already achieved by \xxx{}, our future work will explore how sacrificing some accuracy (e.g., quantization and early-exit policies~\cite{laskaridis2020spinn_mobicom}) can further enhance its efficiency.

\myparagraph{Future Work}
First, although transformer-based models are not yet commonly deployed directly on mobile devices, their increasing adoption underscores the need for continuous enhancement of \xxx{}. 
Addressing this, future work will focus on expanding support for a more diverse range of local operator types and on developing lightweight synchronization techniques for global operators. 
For instance, for operators like Softmax~\cite{liu2016large}, synchronizing only the intermediate sum results instead of the entire input tensor can significantly reduce synchronization overhead; the feasibility of exploiting such computational characteristics has been validated in FlashAttention~\cite{dao2022flashattention}.
Second, it is important to acknowledge that the system performance of \xxx{} is directly influenced by the solution quality achieved by \xxxschedule{}. 
As achieving global optimality in non-convex optimization problems remains a formidable challenge, future research will concentrate on refining the underlying solving algorithms to enhance solution quality, thereby improving scheduling efficiency and further boosting the overall system performance of \xxx{}.

\section{Conclusion}
\label{sec:conclusion}
In this paper, we have proposed \xxx{}, a high-performance collaborative inference system optimized for MEC networks. 
By leveraging \xxxparallel{} for fine-grained parallel execution and \xxxschedule{} for optimized scheduling, \xxx{} dramatically reduces transmission overhead in MEC environments through computation-communication overlap across local operations.
We envision that the fast and energy-efficient inference of \xxx{} will enable diverse tasks on real-world mobile devices, driving broader adoption of advanced machine learning in mobile applications.

\bibliographystyle{IEEEtran}
\bibliography{references}

\vfill

\end{document}